\documentclass[11pt]{article}
\pdfoutput=1
\usepackage{jheppub}

\usepackage{amssymb,amsmath}
\usepackage{amsfonts}
\usepackage{mathtools}
\usepackage{tikz-cd}
\allowdisplaybreaks

\usepackage{physics}
\usepackage{cancel}
\usepackage{enumitem}
\usepackage{bm}
\usepackage{mathrsfs}

\let\originalleft\left
\let\originalright\right
\renewcommand{\left}{\mathopen{}\mathclose\bgroup\originalleft}
\renewcommand{\right}{\aftergroup\egroup\originalright}

\makeatletter
\newenvironment{equations}[1][]{\subequations\ifx\relax#1\relax\else\label{#1}\fi\align\ignorespaces}{\endalign\ignorespacesafterend\endsubequations}
\def\@spliteq#1{\begin{equation}\begin{split}#1\end{split}\end{equation}}
\def\splitequation{\collect@body\@spliteq}

\makeatother

\usepackage{xspace}
\renewcommand{\d}{\ifmmode\operatorname{d}\!\else\textrm{d}\xspace\fi}

\newcommand{\nn}{\nonumber}
\newcommand{\diff}{\mathrm{d}}

\makeatletter
\gdef\@fpheader{}
\makeatother

\title{One Ring to Rule Them All: A Unified Topological Framework for 4D $\mathcal{N}=1$ Superconformal Anomalies}

\author[a,b]{Camillo Imbimbo,}
\author[a]{Ludovico Porro}
\affiliation[a]{Dipartimento di Fisica, Universit\`a di Genova, Via Dodecaneso 33, 16146 Genoa, Italy}
\affiliation[b]{INFN, Sezione di Genova, Genoa, Italy}

\emailAdd{camillo.imbimbo@ge.infn.it}
\emailAdd{ludovico@vt.edu}

\abstract{
We present a unified topological description of anomalies that generalizes the Chern-Simons formulation of Yang-Mills anomalies to encompass all 4-dimensional $\mathcal{N}=1$ superconformal anomalies. The key innovation is our characterization of anomalies through the constraint ideal in the polynomial ring of generalized curvatures and connections of the underlying symmetry (super)-Lie algebra. 

We demonstrate that anomalies in dimension $d$ are captured by the cohomology $H_\delta(W_{d+2})$ of the generalized BRST operator $\delta$ acting on the fermion number $d+2$ component of the constraint ideal $W_{d+2}$. While Yang-Mills anomalies correspond to invariant Chern curvature polynomials (where $W_{d+2}$ reduces to homogeneous curvature polynomials), the constraint ideal for 4D (super)conformal gravity contains additional polynomials mixing curvatures and connections. This richer structure naturally explains the coexistence of both Chern-type ($a$) and non-Chern-type ($c$) anomalies in (super)conformal theories.

}


\begin{document}
\maketitle

\section{Introduction and summary}

All anomalies are BRST cohomology classes \cite{Becchi:1975nq}. However (super)conformal anoma\-lies  \cite{Bonora:1983ff}, \cite{Bonora:1984pn},\cite{Bonora:1985cq} have historically been understood differently from their Yang-Mills and gravitational counterparts. The latter admit an elegant formulation through Chern invariants and Chern-Simons polynomials \cite{Stora:1976LM},\cite{Langouche:1984gn},\cite{Bardeen:1984pm},\cite{Manes:1985df},\cite{Alvarez-Gaume:1983ihn},  whereas no comparable topological framework existed for (super)conformal anomalies until recently. Our previous work \cite{Imbimbo:2023sph} showed that the 4D $\mathcal{N}=1$ superconformal $a$-anomaly can be expressed as a super-Chern-Simons polynomial. However, the existence of two distinct anomalies  \cite{Bonora:1984pn},\cite{Anselmi:1997am} ($a$ and $c$) in 4D (super)conformal gravity suggests a more fundamental structure awaits discovery.

In this paper, we reveal and systematically develop this structure. We demonstrate that \emph{all} 4D $\mathcal{N}=1$ superconformal anomalies admit a unified description that parallels the Chern-Simons formulation of Yang-Mills anomalies. This requires a natural but nontrivial generalization of Chern invariants ---  the central conceptual advance of our work.

The mathematical core of our generalization lies in the \textbf{constraint ideal} of the underlying Lie (super)algebra. To illustrate this, consider the Stora-Zumino framework \cite{Zumino:1983ew},\cite{Stora:1984},\cite{Manes:1985df} for Yang-Mills anomalies, which involves:
\begin{itemize}
\item The \textbf{generalized connection} $\bm{A} = c + A$, combining ghost $c= c^i\, T_i$ (form degree 0) and gauge field $A =A^i\, T_i$ (form degree 1) both valued in the underlying Yang-Mills Lie algebra, with generators $\{ T^i\}$. $\bm{A}$ has  total fermion degree +1.\footnote{Generalized forms $\bm{\Omega}_n =\sum_{p+q=n} \Omega^{(p)}_q$ of  total degree $n$ are the sum of ordinary forms $\Omega^{(p)}_q$ of different form degrees $p$ and ghost numbers $q$, such that $n=q+p$ is the total fermion number.}  
\item The \textbf{generalized BRST differential} $\delta = s + \dd$, encoding the nilpotent BRST operator $s$ and exterior derivative $\dd$.  Nilpotency $\delta^2=0$ follows from $s^2=\{s,\dd\}= \dd^2=0$ and $\delta$ raises the total fermion number by 1.
\item The \textbf{generalized curvature} $\bm{F} = \delta \bm{A} + \bm{A}^2= \bm{F}^i\, T_i$,  satisfying the Bianchi identity $\delta \bm{F} = -[\bm{A}, \bm{F}]$ and carrying total fermion number +2.
\end{itemize}
Crucially, generalized forms of total degree $n>d$
in $d$ space-time dimensions \textbf{do not} need to  vanish, in contrast with ordinary differential forms. 

We adopt a shifted perspective with respect to the original Stora-Zumino framework:  we take
\begin{equations}
& \delta \,\bm{A} =\bm{F}  - \bm{A}^2\label{intro:deltaAFa}\\
&\delta\, \bm{F} = - [\bm{A}, \bm{F}]\label{intro:deltaAFb}
\end{equations}
 as \emph{definitions} of $\delta\bm{A}$ and $\delta\bm{F}$.  Nilpotency $\delta^2 = 0$ then  follows from the Jacobi identity, independent of the $s+d$ decomposition.

Assigning fermion number +1  to  $\bm{A}$ and  +2 to  $\bm{F}$, we consider the graded space of polynomials $P(\bm{A}, \bm{F})$
 \begin{equation}
 P(\bm{A}, \bm{F}) = \bigoplus_{n\ge0} P_n (\bm{A}, \bm{F})
 \end{equation}
where  $P_n$ is the component of fermion number $n$. 
$\delta$ acts as a coboundary operator on this complex:
 \begin{equation}
 \delta : P_n (\bm{A}, \bm{F}) \to P_{n+1} (\bm{A}, \bm{F})
 \end{equation}
A fundamental fact, valid  for {\it any} Lie (super)algebra,  states that the $\delta$-cohomology $H_\delta(P_n)$ is trivial for  all  $n > 0$.\footnote{Decompose $\delta = \delta_0 + \delta_1$, where $\delta_0 \bm{A} = \bm{F}$, $\delta_0 \bm{F} = 0$, and $\delta_1 \bm{A} = -\bm{A}^2$, $\delta_1 \bm{F} = -[\bm{A}, \bm{F}]$. Since $(\bm{A}, \bm{F})$ form a $\delta_0$-trivial doublet, the cohomology of $\delta$ vanishes by standard filtration arguments.}

 In 4D, anomalies correspond to $\delta$-cohomology classes on generalized forms with total degree $n=5$. However, the relevant functional space for Yang-Mills theory is \textbf{constrained} by the ``horizontality'' condition 
 \begin{equation}
    \bm{F} = F, \label{intro:horizontal}
\end{equation}
 where $F$ is the ordinary 2-form Yang-Mills curvature of ghost number 0.  This leads in 4 dimensions to the cubic constraints in $P(\bm{A}, \bm{F})$:
\begin{equation}
  \bm{F}^i \bm{F}^j \bm{F}^k = 0. \label{intro:cubicYM}
\end{equation}
Let $W^{\text{YM}_4}$ denote the \textbf{ideal}\footnote{An ideal $W$ of a commutative ring is $P$ a sub-ring  of $P$ which is invariant under multiplicative action by elements of $P$: $\forall w\in W, p \in P \Rightarrow   p\cdot w \in W$.}   generated by these \textbf{cubic} curvature monomials.  $W^{\text{YM}_4}$ is $\delta$-invariant:
\begin{equation}
    \delta : W^{\text{YM}_4} \to W^{\text{YM}_4}. \label{W3deltainvariance}
\end{equation}
Hence, the field space relevant for Yang-Mills theory is the {\bf quotient}:
\begin{equation}\label{intro:quotientYM}
 P(\bm{A}, \bm{F})/W^{\text{YM}_4},
 \end{equation} 
and Yang-Mills anomalies are elements of $H_\delta(P_5/W^{\text{YM}_4}_5)$, where $W^{\text{YM}_4}_n$ is the fermion number $n$ component of the ideal $W^{\text{YM}_4}$.  

The triviality of $\delta$ on $P_n$ implies the general lemma\footnote{This isomorphism arises because a $\delta$-closed $w_{n+1} \in W_{n+1}$ satisfies $w_{n+1} = \delta p_n$ for some $p_n \in P_n$, where $[p_n] \in P_n/W_n$ is $\delta$-closed. Changing $w_{n+1}$ by $\delta w_n$ shifts $p_n$ by $w_n$, leaving $[p_n]$ unchanged.}
\begin{equation}
    H_\delta(P_n/W_n) \cong H_\delta(W_{n+1}), \label{intro:HWlemma}
\end{equation}
valid for any $\delta$-invariant ideal $W_n$ of $P_n$.   Thus, 4D Yang-Mills anomalies are classified by $H_\delta(W^{\text{YM}_4}_6)$.\footnote{Similarly, $W^{\text{YM}_d}$  in dimension $d$  is generated by curvature monomials of fermion degree  $d+2$ and anomalies are  classified by $H_\delta(W^{\text{YM}_d}_{d+2})$. } 

For Yang-Mills theory (\ref{intro:HWlemma}) simplifies:  in $d=4$,  for example,  $W^{\text{YM}_4}_5=0$ as all elements in  $P_5$ contain at most 2 curvatures.  Hence  
\begin{equation}\label{intro:kerdeltaequalsH}
H_\delta(W^{\text{YM}_4}_6) = \ker_\delta(W^{\text{YM}_4}_6)
\end{equation} 
Furthermore,  since $\delta$ acts on curvatures $\bm{F}$  in the adjoint of the Lie algebra,  $d=4$  Yang-Mills anomalies  in the fermion number 6 picture reduce to \textbf{gauge-invariant} cubic curvature polynomials. This is the familiar one-to-one correspondence between Yang-Mills anomalies in dimension $d$ and Chern polynomials of degree $d+2$. 

In this paper we demonstrate that all $d=4$, $\mathcal{N}=1$ (super)conformal anomalies fit this same framework:  the same isomorphism  (\ref{intro:HWlemma})  also describes (super)conformal anomalies in 4D but with a richer constraint ideal $W$.  We conjecture that (\ref{intro:HWlemma}) describes anomalies associated to any (super)Lie algebras in any dimension $d$, with $n=d+1$.

The isomorphism  (\ref{intro:HWlemma}) gives rise to two equivalent \textbf{pictures}  of anomalies:  in  one picture anomalies in dimension $d$ are  $\delta$-closed polynomials $p_{d+1}$  of fermion degree $d+1$ modulo  contraints and modulo $\delta$-trivial terms
\begin{equation}\label{intro:picturen}
p_{d+1} \sim  p_{d+1}^\prime = p_{d+1} + w_{d+1} + \delta\, p_{d}\qquad \forall \; p_{d}\in P_{d} , \quad p_{d+1}\in P_{d+1}, \quad  w_{d+1}\in W_{d+1}
\end{equation}
In the second picture anomalies are  $\delta$-closed polynomials $w_{d+2}$ of fermion degree $d+2$ of the vanishing ideal  modulo $\delta$-trivial terms
\begin{equation}\label{intro:picturenplus1}
w_{d+2} \sim  w_{d+2}^\prime = w_{d+2} + \delta\, w_{d+1}\qquad \forall \; w_{d+1}\in W_{d+1},  \;w_{d+2}\in W_{d+2}
\end{equation}
For Yang-Mills theory in dimension $d$, $W^{\text{YM}_d}_{d+1}=0$ and hence the equivalence relation  (\ref{intro:picturenplus1})  trivializes:  in this case the fermion number $d+2$ ``picture''  describes the anomalies invariantly by means of  Chern curvature polynomials of fermion degree $d+2$. In the (super)conformal case $W_5$ is non-trivial.  Hence, in this context, invariant Chern polynomials of degree $d+2$ are generalized to  equivalence classes (\ref{intro:picturenplus1}) of $\delta$-closed polynomials of degree $d+2$ in the vanishing ideal.
In all cases, the $d+2$ picture is more ``invariant''  than the $d+1$ picture and it is usually easier to compute, since it involves the computation of the cohomology of $\delta$ on a polynomial space with no constraints.

Let us  briefly describe the structure of both conformal and  the (super)conformal ideals, leaving the details to the rest of the article.
For the  {\it bosonic} $d=4$  $SO(4,2)$ conformal algebra the generalized connections/curvatures are:
\begin{equations}
 &  \bm{A} = \bm{e}^a\, P_a +\bm{\omega}^{ab}\, J_{ab} + \bm{f}^a\,K_a+  \bm{b}\, D\\
 &  \bm{F} = \bm{T}^a\, P_a + \tilde{\bm{R}}^{ab}\, J_{ab}+ \tilde{\bm{T}}^a\, K_a+ \bm{\Tilde{F}}_W \,D
 \end{equations}
where $\{P_a, J_{ab}, K_a, D\}$ are the generators of translations, Lorentz rotations, special conformal transformation and dilations, respectively. The $\delta$-invariant ideal $W^{\text{conf}}$ which characterizes  conformal gravity   is \textbf{generated} 
not only by cubic curvature monomials  but also by connection-dependent constraints:
\begin{equation}\label{intro:conformalW}
W^{\text{conf}}:\quad  \{  \bm{F}^i \bm{F}^j \bm{F}^k, \; \bm{T}^a,  \bm{\Tilde{F}}_W , \bm{e}_a\, \tilde{\bm{T}}^{a},  \bm{e}_a \,\tilde{\bm{R}}^{ab}, \epsilon_{abcd}\,\bm{e}^b  \tilde{\bm{R}}^{cd}, \epsilon_{abcd}\, \bm{e}^b\, \bm{e}^c\,  \tilde{\bm{T}}^{d}\}
 \end{equation}
As we will review in the main text of the article,  these extra constraints arise by the need to incorporate diffeomorphisms into  the definition of the BRST operator.  Hence  the relevant field space for conformal gravity is the {\bf quotient}\begin{equation}\label{intro:quotientCG}
 P(\bm{A}, \bm{F})/W^{\text{conf}}
 \end{equation} 
and its anomalies admit both the 5 and the 6 degree picture:
\begin{equation}\label{intro:Hisomorphism}H_\delta(P_5/W^{\text{conf}}_5)  \equiv  H_\delta(W^{\text{conf}}_{6}) 
\end{equation}
This expanded ideal has two crucial consequences:
\begin{itemize}
\item $W^{\text{conf}}_5 \neq 0$, meaning anomalies  in the fermion number 6 picture correspond to equivalence classes of polynomials rather than single representatives;
\item The cohomology $H_\delta(W^{\text{conf}}_6)$ contains both the unique gauge-invariant cubic curvature polynomial ($a$-anomaly) and a (equivalence class of) non-invariant polynomial of both curvatures and connections (related to the $c$-anomaly) \cite{Bonora:1983ff},\cite{Deser:1993yx}.
\end{itemize}
 
The situation for the superconformal algebra  $\mathfrak{su}(2,2|1)$  (\ref{intro:su221AFa}-\ref{intro:su221AFb}) is conceptually similar  although technically more intricate.  $\mathfrak{su}(2,2|1)$   generalized connections and curvatures are 
\begin{equations}
     & \bm{A}^i=\;\{ \bm{e}^a,\;\bm{\omega}^{ab},\; \bm{f}^a, \;\bm{b},\;\bm{a},\;\bm{\psi}^\alpha,\;\bm{\Tilde{\psi}}^\alpha \}
      \label{intro:su221AFa}\\
       & \bm{F}^i=\;\{ \bm{T}^a,\;\tilde{\bm{R}}^{ab},\; \bm{\Tilde{T}}^a,\;\bm{\Tilde{F}}_W ,\;\bm{\Tilde{F}}_R,\;\bm{\rho}^\alpha,\;\bm{\Tilde{\rho}}^\alpha \},
       \label{intro:su221AFb}
\end{equations}
corresponding to the (graded) generators  $ T_i=\;\{P_a,\;J_{ab}, K_a,\;D,\; R,\; Q_\alpha,\; S_\alpha \}$, 
relative to translations, Lorentz rotations, special conformal transformations,  dilations, R-symmetry, supersymmetry and special conformal supersymmetry.  The nilpotent $\delta$ defined by the universal formulas (\ref{intro:deltaAFa}-\ref{intro:deltaAFb}) is cohomologically trivial just  as in any (super)Lie algebra.\footnote{It is important to keep in mind however that the relation between the generalized $\delta$ and the BRST operator  $s$  for both conformal and superconformal gravity has to be modified with respect to the Yang-Mills theory \cite{Imbimbo:2009dy},\cite{Imbimbo:2018duh},\cite{Imbimbo:2023sph}, as it will reviewed in Section \ref{sec:confgravity}.}

The superconformal ideal  $W^{\text{super-conf}}$  originates from the constraints of superconformal gravity,    described in the literature as relations among  {\it ordinary} curvatures of superconformal gravity \cite{Kaku:1977rk},\cite{Kaku:1978nz},\cite{VanNieuwenhuizen:1981ae},\cite{Fradkin:1985am}\,\cite{Baulieu:1986ab},\cite{vanNieuwenhuizen:2004rh}.  We show in this paper that  these can all be  recast as polynomial relations among generalized curvatures and connections, giving rise to an ideal in $P(\bm{A}, \bm{F})$ generated by  
  \begin{equation}\label{intro:superconformalW}
W^{\text{super-conf}} :\qquad   \{ \bm{T}^a,\;  \bm{e}^a\, \Gamma_a\, \bm{\rho}, \;  \epsilon_{abcd}\,\bm{e}^b\,\bm{\Tilde{R}}^{cd}+2\,\bm{e}_a\,\bm{\Tilde{F}}_R-4\,i\,\bm{\Bar{\psi}}\,\Gamma_{a}\,\Gamma_{_5}\,\bm{\rho} \}
 \end{equation}
Unlike the bosonic case, not all cubic curvature polynomials vanish in $W^{\text{super-conf}}_6$. Nevertheless, the unique $\mathfrak{su}(2,2|1)$ invariant  cubic  curvature polynomial (the $a$-anomaly) does sit in $W^{\text{super-conf}}$, while a second cohomology class (the $c$-anomaly) emerges from non-gauge-invariant  but $\delta$-closed polynomials.

Summarizing, our framework unifies anomaly descriptions through:
\begin{itemize}
\item \textbf{Universal Structure}: The polynomial ring $P(\bm{A}, \bm{F})$ with trivial $\delta$-cohomology
\item \textbf{Theory-Specific Constraints}: Each theory defines a $\delta$-invariant ideal $W(\bm{A}, \bm{F})$
\item \textbf{Anomaly Classification}: $H_\delta(P_{d+1}/W_{d+1}) \cong H_\delta(W_{d+2})$ in $d$ dimensions
\end{itemize}

For Yang-Mills, $W$ is generated by curvature constraints, yielding Chern invariants. For (super)conformal gravity, $W$'s additional connection-dependent constraints naturally explain both $a$- and $c$-type anomalies within a single topological framework.

The rest of this article is organized as follows.  

In Section \ref{sec:confgravity} we review  the formulation of bosonic conformal gravity in the Stora-Zumino  formalism, we describe the corresponding ideal $W^{\text{conf}}$ and  compute the anomaly cohomologies $H_\delta(P_5/W^{\text{conf}}_5)  = H_\delta(W^{\text{conf}}_{6})$ explicitly, recovering both $a$ and $c$ conformal anomalies. 

In Section \ref{sec:scg} we review  the formulation of superconformal gravity in the generalized  BRST formalism and  
we identify the polynomial relations involving both generalized curvatures and connection which generate the superconformal  ideal $W^{\text{super-conf}}$. The components of ghost number 0 of the ideal generators  reproduce the constraints for ordinary curvatures of conformal supergravity.  Moreover, we will see that  the components at higher ghost number of the generators completely determine the non-horizontal components of the generalized curvatures. We then recover both $a$ and $c$ superconformal anomalies in the degree 6 picture by computing  $H_\delta(W^{\text{super-conf}}_{6})$.  By using 
the isomorphism (\ref{intro:HWlemma}) we are able to write down explictly the generalized Chern-Simons polynomials of degree 5 describing the general superconformal anomaly to all order in the fermionic fields: this allows us to correct some errors which appeared in the recent literature regarding the part of the supersymmetry anomaly which captures anomalous correlators with 4 currents (2 supercurrents and 2 stress energy tensors).

\section{Bosonic conformal gravity}\label{sec:confgravity}

\subsection{BRST structure}

Four-dimensional bosonic conformal gravity arises as the gauge theory associated with the $SO(4,2)$ conformal algebra. The fundamental objects are the generalized connections and curvatures:
\begin{equations}\label{seccg:generalizedcc}
 &  \bm{A} = \bm{e}^a\, P_a +\bm{\omega}^{ab}\, J_{ab} + \bm{f}^a\,K_a+  \bm{b}\, D\\
 &  \bm{F} = \bm{T}^a\, P_a + \tilde{\bm{R}}^{ab}\, J_{ab}+ \tilde{\bm{T}}^a\, K_a+ \bm{\Tilde{F}}_W \,D
 \end{equations}where $\{P_a, J_{ab}, K_a, D\}$ are, respectively,  the generators of translations, Lorentz rotations, special conformal transformation and dilations. The BRST structure follows from Eqs. (\ref{intro:deltaAFa}-\ref{intro:deltaAFb}), specializing to:
\begin{subequations}\label{deltaconformalgravity} 
    \begin{align}
     & \delta \bm{b}= \bm{\Tilde{F}}_W-2\,\bm{e}^a\,\bm{f}_a \\
       & \delta \bm{e}^a= \bm{T}^a-\bm{b}\, \bm{e}^a -\boldsymbol{\omega}^{ab}\,\bm{e}_b \\
       & \delta \bm{f}^a=\bm{\Tilde{T}}^a+\bm{b}\,\bm{f}^a-\boldsymbol{\omega}^{ab}\,\bm{f}_b \\
       & \delta \boldsymbol{\omega}^{ab}= \bm{\Tilde{R}}^{ab}-\boldsymbol{\omega}^{ac}\,\boldsymbol{\omega}^{b}\,_c-2\,\bm{e}^{[a}\,\bm{f}^{b]}\\
       & \delta \bm{\Tilde{F}}_W= -2\,\bm{e}^a\,\bm{\Tilde{T}}_a+2\,f^a\,\bm{T}_a \label{deltaFW} \\
        &\delta \bm{T}^a= \bm{e}^a\,\bm{\Tilde{F}}_W+\bm{\Tilde{R}}^{ab}\,\bm{e}_b-\bm{b}\,\bm{T}^a-\boldsymbol{\omega}^{ab}\,\bm{T}_b \label{deltaT} \\
       & \delta \bm{\Tilde{T}}^a= -\bm{f}^a\,\bm{\tilde{F}_W}+\bm{\Tilde{R}}^{ab}\,\bm{f}_b+\bm{b}\,\bm{\Tilde{T}}^a- \boldsymbol{\omega}^{ab}\,\bm{\Tilde{T}}_b \\
       & \delta \bm{\Tilde{R}}^{ab}= \boldsymbol{\omega}^{[ac}\,\bm{\Tilde{R}}^{b]}\,_c-2\,\bm{e}^{[a}\,\bm{\Tilde{T}}^{b]}-2\,\bm{f}^{[a}\,\bm{T}^{b]} \label{deltaRtilde}.
 \end{align}
\end{subequations}
The Stora-Zumino formalism requires modifications for gravity theories compared to Yang-Mills theory \cite{Imbimbo:2023sph}.  The key difference stems from the need to introduce an anticommuting ghost vector field $\xi^\mu$ for diffeomorphisms. The BRST operator $s$ acts  on tensor  fields $\phi$ via the Lie derivative $\mathcal{L}_\xi$\footnote{The minus sign in front of the Lie derivative is traditional in a certain stream of literature.}
\begin{equation}\label{seccg:BRSTdiffeos}
s\, \phi = -\mathcal{L}_\xi\, \phi + \mathrm{other\; gauge\; transformations},
\end{equation} 
and  on the ghost $\xi^\mu$ itself as:
\begin{equation}\label{secone:sxi}
s\, \xi^\mu = - \tfrac 1 2 \,  \mathcal{L}_\xi\,\xi^\mu .
\end{equation}
To avoid redundancy, one must not introduce ghosts associated to  translations $P^a$  and thus must impose the ``horizontality'' condition on the generalized vierbein connection:
\begin{equation}\label{seccg:horizontale}
\bm{e^a} = e^a
\end{equation}
This modifies the relationship between the generalized $\delta$ and BRST operator \cite{Stora:1984}:
\begin{equation}\label{secgc:deltaconformal}
\delta = s + \mathcal{L}_\xi + \dd
\end{equation}

\subsection{Constraint Ideal of Conformal Gravity}

Horizontality of the vierbein (\ref{seccg:horizontale}) implies horizontality of conformal  generalized curvatures 
$$\{\bm{T}^a,   \bm{\Tilde{R}}^{ab}, \bm{\Tilde{T}}^a,\bm{\Tilde{F}}_W\}$$
as in the Yang-Mills case.  Hence the constraint ideal $W^{\text{conf}}$ for 4D conformal gravity contains cubic homogeneous
curvatures polynomials $W^{\text{YM}_4}$. However conformal gravity is characterized by additional constraints. One of them is the  torsion constraint 
\begin{equation}\label{seccg:torsion}
\bm{T}^a =0
\end{equation}
expressing the spin connection in terms of the vierbein.   Two additional constraints are derived by successive $\delta$-actions
\begin{equation}\label{seccg:FWBianchi}
\bm{\tilde{F}}_W= 0 \qquad \bm{e}_a \, \bm{\Tilde{R}}^{ab} =0
\end{equation}
 The first of this constraints eliminates $\bm{b}$ as a propagating degree of freedom.  Acting on (\ref{seccg:FWBianchi})  with  $\delta$  and  using  Eqs. (\ref{deltaFW})-(\ref{deltaRtilde}) one arrives to  a constraint involving $\tilde{\bm{T}}^{a}$
 \begin{equation}\label{seccg:Ttildeone}
\bm{e}_a\,  \tilde{\bm{T}}^{a}=0
\end{equation}
To describe the  gravitational theory one more constraint --- the ``Ricci  constraint'' --- is needed which eliminates $\bm{f}^a$ as an independent fields: 
\begin{equation}\label{seccg:Ricci}
\epsilon_{abcd} \,\bm{e}^b \,  \bm{\Tilde{R}}^{cd} =0
\end{equation}
$\delta$-invariance of the Ricci constraint together with Eq. (\ref{deltaRtilde}) imply a second constraint on $\tilde{\bm{T}}^{a}$
\begin{equation}\label{seccg:Ttildetwo}
\epsilon_{abcd} \,\bm{e}^b \,  \bm{e}^c\,\bm{\Tilde{T}}^{d} =0
\end{equation}
The generators of the $\delta$ invariant ideal $W^{\text{conf}}$ are\footnote{The horizontality condition on $\bm{e}$ imposes additional constraints, including the vanishing of monomials containing five or more $\bm{e}$ fields. However, these do not affect our analysis, as such terms do not appear in the sector of $W_6$ and $W_5$ considered in this work, as explained in the following section.}
\begin{equation}\label{seccg:conformalW}
 \{  \bm{F}^i \bm{F}^j \bm{F}^k, \; \bm{T}^a,  \bm{\Tilde{F}}_W , \bm{e}_a\, \Tilde{\bm{T}}^{a},  \bm{e}_a \,\tilde{\bm{R}}^{ab}, \epsilon_{abcd}\,\bm{e}^b \, \tilde{\bm{R}}^{cd}, \epsilon_{abcd}\, \bm{e}^b\, \bm{e}^c\,  \tilde{\bm{T}}^{d}\}
 \end{equation}
 
 \subsection{Cohomological Analysis}

We compute anomalies in the fermion number 6 picture, evaluating  $H_\delta(W_6)$. $H_\delta(W_6)$ is non-trivial only in the sector of $W_6$ which satisfies the conditions \cite{Imbimbo:2023sph}:
\begin{itemize}
\item no dependence on  the Lorentz connection $\bm{\omega}^{ab}$;
\item zero Weyl charge;\footnote{In accordance with Eqs. (\ref{deltaconformalgravity}),  $\bm{e}^a$ and $\bm{T}^a$ have Weyl charge +1, $\bm{f}^a$ and $\Tilde{\bm{T}}^{a}$ Weyl charge -1, and all the other connections and curvatures Weyl charge 0.}
\end{itemize}
Moreover $\delta$ commutes with spatial parity, hence one can consider its cohomology either on  the even or the odd spatial parity subspaces $W_6^\pm$,  independently. We will focus on the even parity sector, since even parity Weyl anomalies of non-chiral conformal theories are  usually the ones of physical interest.\footnote{Odd parity conformal (non-supersymmetric) anomalies have been investigated in  \cite{Bonora:2014qla},\cite{Bonora:2015nqa},\cite{Bonora:2017gzz},\cite{Abdallah:2023cdw}.} Even spatial parity densities are described by top-forms of parity -1, if  we assign parity -1 to the Levi-Civita tensor $\epsilon_{abcd}$.  To avoid proliferation of symbols we will  denote the sectors of $W$ independent of the Lorentz connection, of zero Weyl charge and even spatial parity by the same letters $W_{6}$ and $W_{5}$.

It will be useful to introduce a second grading in $P(\bm{A},\bm{F})$, beyond the fermion number:  this grading counts ``the number of fields'', that is it assigns values +1 to both $\bm{A}$ and $\bm{F}$.  Since  $\bm{A}$ carries fermion degree +1 and $\bm{F}$ fermion degree +2,  polynomials of  total fermion degree $ 2 \,n$ decompose as
\begin{equation}
P_{n}  =\bigoplus_{N=[\frac{n+1}{2}]}^{N=n} P_{n; N}
\end{equation}
where $P_{n;  N}$ is the space of polynomials of fermion number $n$ and number of fields $N$.  
Correspondingly, one decomposes $\delta$ as the sum of $\delta_0$ of charge $N=0$ and $\delta_1$ of charge $N=1$:
\begin{equation}
\delta = \delta_0 + \delta_1
\end{equation}
where 
\begin{equations}
 &\delta_0 \bm{A} = \bm{F} \qquad \delta_0 \bm{F} = 0\nn\\
 & \delta_1 \bm{A} = -\bm{A}^2\quad \delta_1 \bm{F} = -[\bm{A}, \bm{F}]
 \end{equations}
and
 \begin{equation}
 \delta_0:\; P_{n,N} \to P_{n+1,N} \qquad  \delta_1: \;P_{n,N} \to P_{n+1,N+1}
 \end{equation}
 Moreover 
 \begin{equation}
 \delta_0^2= \delta_1^2 = \{\delta_0, \delta_1\} =0
 \end{equation}
$W_6$  decomposes as
\begin{equation}
W_6 = W_{6,3}\oplus W_{6,4}\oplus W_{6,5}
\end{equation}
where $W_{6,N}$ are the components of $W_6$ with fixed number of fields $N$, which one can explicitly compute
\begin{equations}\label{seccg:W6N}
  &W_{6,3}=A_1\, \epsilon_{abcd}\,\bm{\tilde{R}}^{cd}\,\bm{T}^a\,\bm{\tilde{T}}^b+  
 A_2 \,  \epsilon_{abcd}\, \bm{\Tilde{F}}_W \, \bm{\tilde{R}}^{ab}\,  \bm{\tilde{R}}^{cd}\\
  & W_{6,4}= B_{1}\, \epsilon_{abcd}\,\bm{f}^c\bm{e}^d\,\bm{\Tilde{F}}_W\,\bm{\tilde{R}}^{ab} + 
 B_{2} \,\epsilon_{abce} \,\bm{f}^e\,\bm{e}^d\,\bm{\tilde{R}}^{ab}\,\bm{\tilde{R}}^{c}_{\;d} + 
 B_{3} \,\epsilon_{abcd}\,\bm{f}^d\,\bm{b}\,\bm{\tilde{R}}^{bc}\, \bm{T}^a  +\nn\\ 
&\qquad + B_{4}\,\epsilon_{abcd}\,\bm{e}^d\,\bm{b}\,\bm{\tilde{R}}^{bc}\,\bm{\tilde{T}}^a + 
 B_{5}\,\epsilon_{abcd}\,\bm{f}^c\,\bm{e}^d\,\bm{T}^a\,\bm{\tilde{T}}^b \\
  & W_{6,5}= C_{1}\,\epsilon_{abcd} \, \bm{f}^a\,\bm{f}^b\,\bm{e}^c\,\bm{e}^d\,\bm{\Tilde{F}}_W + 
 C_{2}\, \epsilon_{acde} \, \bm{f}^c\,\bm{f}^d\,\bm{e}^b\,\bm{e}^e\,\bm{\tilde{R}}^{a}_{\;b} +\nn\\ 
 & \qquad +
  C_{3}\, \epsilon_{abcd}\,\bm{f}^b\,\bm{f}^c\,\bm{e}^d\,\bm{b}\,\bm{T}^a + 
 C_{4}\, \epsilon_{abcd}\, \bm{f}^b\,\bm{e}^c\,\bm{e}^d\,\bm{b}\,\bm{\tilde{T}}^a\\
 &W_{6,6}=0
\end{equations}
Hence 
\begin{equation}
\dim W_6 = 11
\end{equation}
For $W_5$ we obtain 
 \begin{equation}
 W_5 = W_{5,3}\oplus W_{5,4}
 \end{equation}
 where
\begin{equations}\label{seccg:W5N}
  & W_{5,3}=a_1\, \epsilon_{abcd}\,\bm{f}^a\,\bm{\tilde{R}}^{cd}\,\bm{T}^b  + 
 a_2\,
    \epsilon_{abcd}\,\bm{e}^a\,\bm{\tilde{R}}^{cd}\,\bm{\tilde{T}}^b\\
&W_{5,4} =b_1\,\epsilon_{abcd} \,\bm{f}^b\,\bm{f}^c\,\bm{e}^d\,\bm{T}^a 
+ b_2 \, \epsilon_{abcd} \,   \bm{f}^b\,\bm{e}^c\bm{e}^d\,\bm{\tilde{T}}^a+ 
 b_3\,\epsilon_{abcd}\,\bm{f}^c\,\bm{e}^d\,\bm{b}\,\bm{\tilde{R}}^{ab} \\
 &W_{5,5}=0
\end{equations}
Hence 
\begin{equation}
\dim W_5 = 5
\end{equation}
The equations for the kernel of $\delta$ at fermion degree 6
\begin{equation}\label{seccg:kerneldelta}
\delta\, W_6 =0
\end{equation}
are equivalent to the ``descent equations'' 
\begin{equations}\label{seccg:kerneldeltadescent}
&\delta_0 \, W_{6,3}=0\\
&\delta_1 \, W_{6,3} + \delta_0\, W_{6,4}=0\\
&\delta_1 \, W_{6,4}+ \delta_0\, W_{6,5}=0\\
&\delta_1 \, W_{6,5}=0
\end{equations}
 Inserting  Eqs. (\ref{seccg:W6N}) into  Eqs. (\ref{seccg:kerneldeltadescent})  one obtains
\begin{equations}
   & W_{6,3}^{\text{closed}}=A_1\,\epsilon_{abcd}\,\bm{\tilde{F}_W}\,\bm{\tilde{R}}^{ab}\,\bm{\tilde{R}}^{cd}+A_2\,\epsilon_{abcd}\,\bm{\tilde{R}}^{cd}\,\bm{T}^a\,\bm{\tilde{T}}^b\nn\\
& W_{6,4}^{\text{closed}}= B_{1}\,\epsilon_{abcd}\,\bm{f}^c\, \bm{e}^d\, \bm{\tilde{F}_W}\,\bm{\tilde{R}}^{ab}+(8\,A_1-A_2)\,\epsilon_{abce}\,\bm{f}^e\,\bm{e}^d\, \bm{\tilde{R}}^{ab}\,\bm{\tilde{R}}^{c}_{\;d}+\\  
&\qquad  +(8\,A_1-A_2+B_1)\, (\epsilon_{abcd}\,\bm{f}^d\,\bm{b}\,\bm{\tilde{R}}^{cd}\,\bm{T}^a + \epsilon_{abcd}\, \bm{e}^d \,\bm{b}\,\bm{\tilde{R}}^{cd}\,\bm{\tilde{T}}^a) \\ 
 &\qquad  + B_{5} \,\epsilon_{abcd}\,\bm{f}^c\,\bm{e}^d \,\bm{T}^a\,\bm{\tilde{T}}^b\\
& W_{6,5}^{\text{closed}}=  C_{1}\,\epsilon_{abcd}\,\bm{f}^a\,\bm{f}_b\, \bm{e}^{c}\, \bm{e}^d\, \bm{\tilde{F}_W}+\frac{1}{2}(-4\,B_1+B_5)\,\epsilon_{abce}\,\bm{f}^c\, \bm{f}^d\,\bm{e}^b\,\bm{e}^e\, \bm{\tilde{R}}^{ab}\\ 
&\qquad +(4\,B_1-B_5+C_1)\,\epsilon_{abcd}\,(\bm{f}^b\,\bm{f}^c\,\bm{e}^d\,\bm{b}\,\bm{T}^a  +\bm{f}^b\,\bm{e}^c\, \bm{e}^d\, \bm{b}\,\bm{\tilde{T}}^a) 
  \end{equations}
  The dimension of  $\ker_\delta(W_6)$ is therefore 
  \begin{equation}
  \dim \ker_\delta(W_6) = 5
  \end{equation}
  One can  similarly verify that 
    \begin{equation}
  \dim \ker_\delta(W_5) = 2 \Rightarrow \dim \Im_\delta(W_5) = 3
  \end{equation}
From this we conclude that the  cohomology $H_\delta(W_6)$ has dimension 2
  \begin{equation}
    \dim H_\delta(W_6)=   \dim \ker_\delta(W_6)-\dim \Im_\delta(W_5) = 2
  \end{equation}

\subsection{Anomaly Representatives}

$H_\delta(W_6)$ admits 2 natural linear independent representatives.  To characterize them let us first observe that  since all  $\delta_1$-closed  $W_{6,5}^{\text{closed}}$ are $\delta_1$-exact 
\begin{equation}
W_{6,5}^{\text{closed}} = \delta_1 W_{5,4}
\end{equation}
 \emph{both} classes admit representatives  with vanishing $W_{6,5}$ components:
  \begin{equation}\label{seccg:W65gauge}
  W_{6,5}^{\text{closed}}  =0
  \end{equation}
 A natural basis of  $H_\delta(W_6)$  is composed of two classes $A$ and $B$ having \emph{homogenous} representatives $W_{6,N}$ with definite field numbers: 
  \begin{equations} \label{seccg:HAelement}
&  W_{6}^{(A)} = W^{(A)}_{6,3}= -\frac{1}{16\, \pi^2}\,\epsilon_{abcd}\, \bigl(\bm{\Tilde{F}}_W\,\bm{\Tilde{R}}^{ab}\,\bm{\Tilde{R}}^{cd}+8\, \bm{T}^a\,\bm{\Tilde{T}}^b\,\bm{\Tilde{R}}^{cd}\bigr)\\
&  W_{6}^{(B)} = W^{(B)}_{6,4}=\frac{1}{2\, \pi^2}\,\epsilon_{abcd}\,\bigl(\bm{f}^d\,\bm{b}\,\bm{\Tilde{R}}^{bc}\,\bm{T}^a +\bm{e}^d\,\bm{b}\,\bm{\Tilde{R}}^{bc}\,\bm{\Tilde{T}}^a+ \nn \\ & \qquad \qquad \qquad\qquad\qquad  + \bm{f}^c\,\bm{e}^d\,\bm{\Tilde{F}}_W\,\bm{\Tilde{R}}^{ab}+4\,\bm{f}^c\,\bm{e}^d\,\bm{T}^a\,\bm{\Tilde{T}}^b \bigr)
\end{equations}
The cohomology representative $W^{(A)}_{6,3}$ contains only curvatures and satifies  the shortened descent equations
\begin{equation}\label{seccg:kerneldeltadescentshort}
\delta_0 \, W^{(A)}_{6,3}=0\qquad \delta_1 \, W^{(A)}_{6,3}=0
\end{equation}
$\delta_1$ acts on curvatures as the adjoint with parameter $\bm{A}$. Hence the representative $W^{(A)}_{6,3}$ is  an \textbf{invariant polynomial}  of the curvatures --- i.e.  it is  the (unique) Chern polynomial of  the $SO(4,2)$ algebra.
It turns out that $W^{(A)}_{6,3}$  is proportional to the  so-called  $a$-conformal anomaly 
\begin{equation}
H_{6}^{(a)} = a \, W^{(A)}_{6,3}
\end{equation}
The  class described by $W^{(B)}_{6,4}$  does not admit a representative of Chern-type containing only curvatures: we might think of it as a generalization of the familiar Chern polynomials.  The  so-called $c$-conformal anomaly    $H_6^{(c)}$ is related to  the  $A$ and $B$ classes as follows
\begin{equation}
	\label{c class conformal}
H_6^{(c)} =  c\, ( W^{(B)}_{6,4}-W^{(A)}_{6,3})
\end{equation}
This relation shows that $W^{(B)}_{6,4}$  describes the anomaly  of conformal theories with $a=c$  
\begin{equation}\label{seccg:Bacrelation}
H_{6,4}^{(B)}=c\, W^{(B)}_{6,4} = (H_{6}^{(a)} + H_{6}^{(c)})\big|_{a=c} 
\end{equation}

Let us turn to the degree 5 ``picture'' of conformal anomalies.  The  $a$-anomaly
\begin{equation}
H_{6}^{(a)} = a\,  W^{(A)}_{6,3} = \delta\, Q^{(a)}_5
 \end{equation}
determines,  up  to $\delta$-exact terms, the ordinary $SO(4,2)$ Chern-Simons polynomial:
 \begin{equation}
 \label{A type conformal Chern-Simons}
    Q_5^{(a)}=-\frac{a}{16\pi^2}\, \epsilon_{abcd}\,\bigl(\bm{b}\,\bm{\Tilde{R}}^{ab}\,\bm{\Tilde{R}}^{cd}+16\,\bm{b}\,\bm{f}^a\,\bm{f}^b\,\bm{e}^c\,\bm{e}^d\bigr)
\end{equation}
The  $c$-anomaly
\begin{equation}
H_{6}^{(c)} = \delta\, Q^{(c)}_5
 \end{equation}
defines a degree 5 class which has the property of admitting a homogeneous representative $Q^{(c)}_{5,3}$
\begin{equation}
	\label{c_type conformal Chern-Simons}
Q_5^{(c)}= Q_{5,3}^{(c)} = \frac{c}{16\pi^2}\,\epsilon_{abcd}\,\bm{b}\,\bm{\Tilde{R}}^{ab}\,\bm{\Tilde{R}}^{cd},
\end{equation}
The 4-form of ghost number  1,  i.e.  the  \emph{the anomaly density},  corresponding to the representative $Q_{5,3}^{(c)}$, is strictly $s$-invariant ---   not just $s$-invariant modulo $d$.\footnote{Weyl anomaly densities which are strictly BRST invariant are named by Deser and Schwimmer of type $B$ \cite{Deser:1993yx}. It might be worth keeping in mind that the  $A$ and $B$ nomenclature of \cite{Deser:1993yx} refers to  anomaly densities, i.e. to the degree 5 picture. Our $W^{(A)}$ and $W^{(B)}$ are  degree 6 picture anomalies.  While $W^{(A)}_{6,3}$ does correspond to an anomaly density  which is of the  $A$-type in the Deser-Schwimmer sense, $W^{(B)}_{6,4}$  does not correspond to their $B$-type anomaly density.  The  degree 6 picture representative of the $B$-type anomaly density of Deser and Schwimmer is $H_6^{(c)}$.}

The $H_{6,4}^{(B)}$ anomaly
\begin{equation}
H_{6,4}^{(B)} = \delta\, Q^{(a=c)}_5
 \end{equation}
defines  a degree 5 class which has the property of admitting a homogenous representative 
\begin{equation}
\label{B type conformal Chern-Simons}
    Q_5^{(a=c)}=Q^{(a=c)}_{5,5} =-\frac{a}{\pi^2}\, \epsilon_{abcd}\,\bm{f}^a\,\bm{f}^b\,\bm{e}^c\,\bm{e}^d\,\bm{b}
\end{equation}
with
\begin{equation}
Q^{(a=c)}_{5,3}=Q^{(a=c)}_{5,4}=0
\end{equation}
Summarizing, the relation between the $A$ and $B$ basis of the degree 6 picture and the $Q_5^{(a)}$ and $Q_5^{(c)}$ basis of the degree 5 picture is  described by the following diagrams:
	\begin{equations}
			 &Q_{5}^{(a)}\xlongrightarrow{\quad\delta\quad}  H_{6,3}^{(A)} \nn \\%
			&Q_{5}^{(a=c)} \xlongrightarrow{\quad\delta\quad}  H_{6,4}^{(B)}\nn\\%
			&Q_{5,3}^{(c)} \xlongrightarrow{\quad\delta\quad}  H_{6,4}^{(B)}-H_{6,3}^{(A)}\nn%
	\end{equations}
Finally, the anomaly density associated to the generic anomaly $Q_5^{(a)}+ Q_5^{(c)}$ is 
\begin{equation}\label{seccg:acanomalies}
	\sigma \mathcal{A}=  -\frac{a}{16\pi^2}\,\sigma \, E_4 + \frac{c}{16\pi^2}\,\sigma \, \epsilon_{\mu\nu\rho\sigma}\,W^{\mu\nu\alpha\beta}\,W^{\rho\sigma}\,_{\alpha\beta}  
\end{equation}
where
\begin{align}
	& E_4= R^{\alpha\beta\gamma\delta}R_{\alpha\beta\gamma\delta}-4R^{\alpha\beta}R_{\alpha\beta}+R^2 \\ 
	& S^{\alpha\beta}=\frac{R^{\alpha\beta}}{2}-\frac{1}{12}g^{\alpha\beta}R \\
	& W^{\alpha\beta\gamma\delta}=R^{\alpha\beta\gamma\delta}-S^{[\alpha[\gamma}g^{\beta]\delta]}
\end{align}
are the Euler density in 4D, the Schouten tensor and the Weyl tensor respectively. 

\section{Superconformal gravity}\label{sec:scg}

\subsection{BRST  structure}

4D superconformal gravity is the gauge theory associated to the $\mathfrak{su}(2,2|1)$ super conformal algebra. Let 
\begin{equations}
     & \bm{A}^i=\;\{ \bm{e}^a,\;\bm{\omega}^{ab},\; \bm{f}^a, \;\bm{b},\;\bm{a},\;\bm{\psi}^\alpha,\;\bm{\Tilde{\psi}}^\alpha \}
      \label{secscg:su221AFa}\\
       & \bm{F}^i=\;\{ \bm{T}^a,\;\tilde{\bm{R}}^{ab},\; \bm{\Tilde{T}}^a,\;\bm{\Tilde{F}}_W ,\;\bm{\Tilde{F}}_R,\;\bm{\rho}^\alpha,\;\bm{\Tilde{\rho}}^\alpha \},
       \label{secscg:su221AFb}
\end{equations}
be the generalized connections and curvatures corresponding to the (graded) generators  $ \{T_i\}=\{P_a,\;J_{ab}, \;K_a,\;  D,\;  R,\;  Q_\alpha,\; S_\alpha \}$, 
relative to translations, Lorentz rotations, special conformal transformations,  dilations, R-symmetry, supersymmetry and special conformal supersymmetry.  

The general formulas for $\delta$ in  Eqs. (\ref{intro:deltaAFa}-\ref{intro:deltaAFb}) specialize to the $\mathfrak{su}(2,2|1)$ superconformal algebra as follows\footnote{See appendix \ref{appendixA} for a review of our notation and conventions for the (anti)commutation relations of   $\mathfrak{su}(2,2|1)$.}
\begin{subequations}\label{secscg:deltasuper}
	\begin{align}
		& \delta \bm{b}= \bm{\Tilde{F}}_W-2\,\bm{e}^a\,\bm{f}_a +2 \,i \, \bm{\Bar{\psi}}\,\bm{\Tilde{\psi}}\\
		& \delta \bm{a}= \bm{\tilde{F}}_R+2\, i\, \bm{\Bar{\psi}}\,\Gamma_{_5}\,\bm{\Tilde{\psi}}\\
		& \delta \bm{e}^a= \bm{T}^a-\bm{b}\,\bm{e}^a -\boldsymbol{\omega}^{ab}\,\bm{e}_b -\bm{\Bar{\psi}}\,\Gamma^a\, \bm{\psi} \\
		& \delta \bm{f}^a=\bm{\Tilde{T}}^a+\bm{b}\,\bm{f}^a-\boldsymbol{\omega}^{ab}\,\bm{f}_b +\bm{\Bar{\Tilde{\psi}}}\,\Gamma^a \,\bm{\Tilde{\psi}} \\
		& \delta \boldsymbol{\omega}^{ab}= \bm{\Tilde{R}}^{ab}-\boldsymbol{\omega}^{ac}\,\boldsymbol{\omega}_c\,^{b}-2\bm{e}^{[a}\,\bm{f}^{b]}+2\,i\, \bm{\Bar{\psi}}\,\Gamma^{ab}\,\bm{\Tilde{\psi}}\\
		& \delta \bm{\psi}= \bm{\rho}-\frac{1}{2}\,\bm{b}\,\bm{\psi}-i\, \bm{e}^a \, \Gamma_a\, \bm{\Tilde{\psi}}+\frac{3}{2}i\, \bm{a}\,\Gamma_{_5} \,\bm{\psi}-\frac{1}{4}\,\bm{\omega}_{ab}\,\Gamma^{ab}\,\bm{\psi} \\
		& \delta \bm{\Tilde{\psi}}= \bm{\Tilde{\rho}}+\frac{1}{2}\,\bm{b}\,\bm{\Tilde{\psi}}+i \,\bm{f}^a \,\Gamma_a\, \bm{\psi}-\frac{3}{2}\,i\, \bm{a}\,\Gamma_{_5} \,\bm{\Tilde{\psi}}-\frac{1}{4}\,\bm{\omega}_{ab}\,\Gamma^{ab}\,\bm{\Tilde{\psi}} \\
		& \delta\bm{\Tilde{F}}_W= -2\,\bm{e}^a\,\bm{\Tilde{T}}_a+2\,\bm{f}^a\,\bm{T}_a-2\,i\, \bm{\Bar{\rho}}\,\bm{\Tilde{\psi}}+2\,i\,\bm{\Bar{\Tilde{\rho}}}\,\bm{\psi} \\&\delta\bm{\Tilde{F}_R}=-2\,i\,\bm{\Bar{\rho}}\,\Gamma_{_5}\,\bm{\Tilde{\psi}}+2\,i\,\bm{\Bar{\Tilde{\rho}}}\,\Gamma_{_5}\,\bm{\psi} \\
		&\delta \bm{T}^a= \bm{e}^a\,\bm{\Tilde{F}}_W+\bm{\Tilde{R}}^{ab}\,\bm{e}_b-\bm{b}\,\bm{T}^a-\boldsymbol{\omega}^{ab}\,\bm{T}_b+2\, \bm{\Bar{\rho}}\,\Gamma^a\,\bm{\psi} \\
		& \delta \bm{\Tilde{T}}^a= -\bm{f}^a\,\bm{\Tilde{F}}_W+\bm{\Tilde{R}}^{ab}\,\bm{f}_b+\bm{b}\,\bm{\Tilde{T}}^a- \bm{\omega}^{ab}\,\bm{\Tilde{T}}_b-2\,\bm{\Bar{\Tilde{\rho}}}\,\Gamma^a\,\bm{\Tilde{\psi}} \\
		& \delta \bm{\Tilde{R}}^{ab}= -\boldsymbol{\omega}^{[ac}\,\bm{\Tilde{R}}_c\,^{b]}-2\,\bm{e}^{[a}\,\bm{\Tilde{T}}^{b]}-2\bm{f}^{[a}\,\bm{T}^{b]}-2\,i\,\bm{\Bar{\rho}}\,\Gamma^{ab}\,\bm{\Tilde{\psi}}-2\,i\,\bm{\Bar{\Tilde{\rho}}}\,\Gamma^{ab}\,\bm{\Tilde{\psi}}\\
		& \nn \delta \bm{\rho}= - \frac{1}{2}\,\bm{b}\,\bm{\rho}-i\bm{e}_a\,\Gamma^a\,\bm{\Tilde{\rho}}-\frac{1}{4}\,\bm{\omega}_{ab}\,\Gamma^{ab}\,\bm{\rho}+\frac{3}{2}\,i\, \bm{a}\,\Gamma_{_5}\,\bm{\rho}-\frac{3}{2}\,i\,\Gamma_{_5}\,\bm{\psi}\,\bm{\Tilde{F}_R}+\frac{1}{2}\,\bm{\psi}\,\bm{\Tilde{F}}_W+ \\ & \qquad +\frac{1}{4}\,\Gamma^{ab}\,\bm{\psi}\,\bm{\Tilde{R}}_{ab}+i\,\Gamma^a\,\bm{\Tilde{\psi}}\,\bm{T}_a \\
		& \nn \delta \bm{\Tilde{\rho}}=  \frac{1}{2}\,\bm{b}\,\bm{\Tilde{\rho}}+i\,\bm{f}_a\,\Gamma^a\,\bm{\rho}-\frac{1}{4}\,\bm{\omega}_{ab}\,\Gamma^{ab}\,\bm{\Tilde{\rho}}-\frac{3}{2}\,i\, \bm{a}\,\Gamma_{_5}\,\bm{\Tilde{\rho}}+\frac{3}{2}\,i\,\Gamma_{_5}\,\bm{\Tilde{\psi}}\,\bm{\Tilde{F}_R}-\frac{1}{2}\,\bm{\Tilde{\psi}}\,\bm{\Tilde{F}}_W+ \\ & \qquad +\frac{1}{4}\,\Gamma^{ab}\,\bm{\Tilde{\psi}}\,\bm{\Tilde{R}}_{ab}-i\,\Gamma^a\,\bm{\psi}\,\bm{\Tilde{T}}_a .
	\end{align}
\end{subequations}
As we already emphasized,   $\delta$ defined above   is nilpotent regardless of any  contraints on generalized connections and curvatures. 

The relation between $\delta$ with the BRST operator $s$ is \cite{Baulieu:1986ab},\cite{Imbimbo:2023sph}, 
\begin{equation}\label{secsgc:deltasuperconformal}
\delta = s + \mathcal{L}_\xi + \dd  - i_\gamma
\end{equation}
where $\xi^\mu$ is the anti-commuting ghost for diffeomorphisms acting on ordinary fields as in Eq. (\ref{seccg:BRSTdiffeos}), $i_\gamma$ is the contraction of a form with the ghost number 2 commuting vector field $\gamma^\mu$
\begin{equation}\label{secsgc:gammaghost}
\gamma^\mu \equiv \bar\zeta \,\Gamma^\mu\, \zeta
\end{equation}
and $\zeta$ is the commuting spinorial ghost associated to the supersymmetry generator $Q_\alpha$. 
In the supersymmetric case the action of $s$ on the ghost $\xi^\mu$ is 
\begin{equation}\label{secscg:sxi}
s\, \xi^\mu = - \tfrac 1 2 \,  \mathcal{L}_\xi\,\xi^\mu  +\gamma^\mu
\end{equation}
Nilpotency of $s$ requires
\begin{equation}\label{secscg:sgamma}
s\, \gamma^\mu = -\mathcal{L}_\xi\, \gamma^\mu
\end{equation}
As in the bosonic case  --- to avoid redundancy ---   the generalized vierbein connection must be horizontal
\begin{equation}\label{secscg:horizontale}
\bm{e^a} = e^a
\end{equation}
\subsection{Constraint Ideal of Superconformal Gravity}\label{secscg:superconformalideal}

In the supersymmetric context not all curvatures are horizontal \cite{Baulieu:1986ab}: only the curvatures associated to physical fields are
\begin{equation}\label{secscg:horizontalcurvatures}
  \bm{T}^a = T^a \qquad \bm{\Tilde{F}}_W=  \Tilde{F}_W\qquad \bm{\Tilde{F}}_R=  \Tilde{F}_R\qquad \bm{\rho} = \rho \;. 
\end{equation}
 Curvatures associated to composite gauge fields have a ghost number 1 components proportional to the supersymmetry ghost $\zeta$:
\begin{equation}\label{secscg:nonhorizontalcurvatures}
 \bm{\Tilde{R}}^{ab}= \Tilde{R}^{ab}+ \lambda_J^{ab} \qquad \bm{\Tilde{T}}^a = \Tilde{T}^a + \lambda_K^a \qquad\bm{\Tilde{\rho}}=  \Tilde{\rho}+ \lambda_S 
\end{equation}
with
\begin{equations} 
 & \lambda_J^{ab}=2\, e^c\,\overline{\zeta}\,\Gamma_c\,\rho^{ab},\label{secscg:lambda_0}\\
&\lambda_K^a =-i\,e^c\,\overline{\zeta}\,\Gamma_c\Gamma_b\,\Tilde{\rho}^{\prime ab},\label{secscg:lambda_0c}\\
&\lambda_S=\tfrac{1}{4}\,\Gamma_5\,\Gamma^{mn}\,\Gamma_c\,\zeta\,\Tilde{F}^R_{mn}e^c .\label{secscg:lambda_0b}
\end{equations}
Hence only the cubic monomials of the horizontal curvatures in Eq.  (\ref{secscg:horizontalcurvatures}) are generators of the constraint ideal $W^{\text{super-conf}}$. 

Another peculiarity of  supersymmetry,   is that horizontality of the vierbein Eq. (\ref{secscg:horizontale}) together with nilpotency of $s$ on the diffeomorphisms ghost  --- Eq. (\ref{secscg:sgamma}) ---  \textbf{dictates} the torsion constraint\footnote{In the bosonic case  horizontality of the vierbein implies horizontality of the torsion $\bm{T}^a$, but not its vanishing. Imposing the vanishing of $\bm{T}^a$ is part of the definition of the conformal ideal which distinguishes it from the   $SO(4,2)$  Yang-Mills  ideal. }
\begin{equation}\label{secsgc:torsionconstraint}
    \bm{T}^a=0
\end{equation}
The $\delta$-variation of the torsion constraint leads to a supersymmetric extension of the Bianchi  identities
\begin{equation}\label{secsgc:superBianchi}
    \bm{e}_a\,\bm{\Tilde{R}}^{ab}- \bm{e}^b \,\bm{\Tilde{F}}_W-2\,\bm{\Bar{\psi}}\,\Gamma^{b}\bm{\rho}=0,
\end{equation}
One must also impose the fermion constraint  \cite{Imbimbo:2023sph}
\begin{equation}\label{secsgc:rhoconstraint}
    \bm{e}_a\,\Gamma^a\,\bm{\rho}=0
\end{equation}
which, upon action with $\delta$,  produces the supersymmetric extension of the Ricci constraint in Eq. (\ref{seccg:Ricci})
\begin{equation}
    \epsilon_{abcd}\,\bm{e}^b\,\bm{\Tilde{R}}^{cd}+2\,\bm{e}_a\,\bm{\Tilde{F}}_R-4\,i\,\bm{\Bar{\psi}}\,\Gamma_{a}\,\Gamma_{_5}\,\bm{\rho}=0
\end{equation}
The $\delta$-variation of the super-Ricci constraint gives the supersymmetric extension of (\ref{seccg:Ttildetwo})
\begin{equation}\label{secsgc:Tttildeconstraint}
\epsilon^{a}\,_{bcd}\,\bm{e}^b\,\bm{e}^c\,\bm{\Tilde{T}}^d+2\,\bm{e}^a\,\bm{\Bar{\psi}}\,\Gamma_{_5}\,\bm{\Tilde{\rho}}+2\,\bm{e}_b\,\bm{\Bar{\psi}}\,\Gamma^{ab}\,\Gamma_5\,\bm{\Tilde{\rho}}-2\,\bm{\Bar{\psi}}\,\Gamma^a\,\bm{\psi}\,\bm{\Tilde{F}}_R=0.
\end{equation}

The super-con\-for\-mal ideal $W^{\text{super-conf}}$ is generated by the polynomial relations in Eqs.  (\ref{secsgc:torsionconstraint}-\ref{secsgc:Tttildeconstraint}).  The components of ghost number 0 of these polynomials constraints  reproduce all the constraints for ordinary curvatures of conformal supergravity. However, the same relations have also  non-trivial components at higher ghost number.  It is quite remarkable that these higher-ghost number constraints  are equivalent to Eq. (\ref{secscg:lambda_0}-\ref{secscg:lambda_0c}), that is they completely determine the non-horizontal components of the curvatures. In other words superconformal gravity curvature constraints and non-horizontal components of the non-physical curvatures are both captured by the polynomial relations among generalized curvatures and connections  which define $W^{\text{super-conf}}$.

To compute anomalies we need to determine the components at ghost number 6 and 5 of $W^{\text{super-conf}}$. As in the bosonic case we can restrict ourselves to the  sector of  Weyl charge 0 and even parity which does not contain the  Lorentz connection.  We computed the dimension (of this sector of) of $W^{\text{super-conf}}_6$ to be\footnote{We performed the relative computations with the help of \textsc{FieldsX} \cite{Frob:2020gdh}.}
 \begin{equation}
\dim W^{\text{super-conf}}_6 = 76
\end{equation}
while the dimension of the vanishing ideal at fermion number 5 is 
\begin{equation}
\dim W^{\text{super-conf}}_5= 19
\end{equation}
Decomposing $W^{\text{super-conf}}_6$  and $W^{\text{super-conf}}_5$  into sectors with fixed number of fields 
\begin{equations}
&W_6 = W_{6,3}\oplus W_{6,4}\oplus W_{6,5}\\
&W_5 = W_{5,4}\oplus W_{5,3}
\end{equations}
we find the vanishing polynomials to be 
\begin{align}\label{secscg:w63}
   & \nn W_{6,3}=A_1 \,\bm{\bar{\tilde{\rho}}}\,\bm{\rho}\,\bm{\Tilde{F}}_R+A_2\,\bm{\Tilde{F}}_R\,\bm{T}^a \,\bm{\tilde{T}}_a+A_3 \, \epsilon_{abcd}\,\bm{\tilde{R}}^{cd} \,\bm{T}^a\,\bm{\tilde{T}}^b+ A_4  \,\bm{\Tilde{F}}_R^3 + 
 A_5\,\bm{\Bar{\tilde{\rho}}}\,\Gamma_5\,\bm{\rho}\,\bm{\Tilde{F}}_W +\\
 & \nn\quad + A_6\, \bm{\Tilde{F}}_R  \,\bm{\Tilde{F}}_W^2 + 
 A_7\, \bm{\bar{\tilde{\rho}}}\,\Gamma^{ab}\,\Gamma_5\,\bm{\rho}\,\bm{\tilde{R}}_{ab} + 
 A_8 \,(\bm{\Tilde{F}}_R\,  \bm{\tilde{R}}_{ab}  \,  \bm{\tilde{R}}^{ab} + 
    \frac{1}{2}\,  \epsilon_{abcd}\, \bm{\Tilde{F}}_W \, \bm{\tilde{R}}^{ab}  \, \bm{\tilde{R}}^{cd}) + \\
    &\quad +
 A_9 \,\bm{\bar{\tilde{\rho}}}\,\Gamma^{a}\,\Gamma_5\,\bm{\tilde{\rho}}\,\bm{T}_a  + 
 A_{10}\, \bm{\bar{\rho}}\,\Gamma^a\,\Gamma_5\,\bm{\rho}\,\bm{\tilde{T}}_a
\end{align}
\begin{align}\label{secscg:w64}
  & \nn W_{6,4}=B_1 \,(\bm{f}^a\bm{e}_a\,\bm{\bar{\tilde{\rho}}}\,\Gamma_5\,\bm{\rho} + 
  \bm{f}^a \,\bm{e}^b \,\bm{\bar{\tilde{\rho}}}\,\Gamma_{ab}\,\Gamma_5\,\bm{\rho}) + 
 B_2 \,\bm{f}^a\,\bm{b}\,\bm{\bar{\rho}}\,\Gamma_a\,\Gamma_5\,\bm{\rho} + \\ 
 & \nn \quad  +B_3\, \bm{\bar{\tilde{\psi}}}\,\Gamma^a\,\bm{\tilde{\psi}}\,\bm{\bar{\rho}}\,\Gamma_a\,\Gamma_5\,\bm{\rho}+   
 B_4\, (\bm{f}^a\bm{e}_a\,\bm{\bar{\tilde{\rho}}}\,\Gamma_5\,\bm{\rho} +\bm{f}^a 
   \,\bm{\bar{\rho}}\,\Gamma_a\,\bm{\psi}\, \bm{\Tilde{F}}_R)   +B_5\, \bm{f}^a\bm{e}_a\,\bm{\Tilde{F}}_R\,\bm{\Tilde{F}}_W + \\ 
   & \nn \quad + B_6\, (\bm{a}\,\bm{\bar{\tilde{\psi}}}\,\Gamma_5\,\bm{\rho}\,\bm{\Tilde{F}}_R + i\, \bm{a}\,\bm{\bar{\tilde{\psi}}}\,\bm{\rho}\, \bm{\Tilde{F}}_W)  + 
 B_7\, (-i \,\bm{b}\,\bm{\bar{\tilde{\psi}}}\,\bm{\rho}\,\bm{\Tilde{F}}_R  + 
    \bm{b}\,\bm{\bar{\tilde{\psi}}}\,\Gamma_5\,\bm{\rho} \,\bm{\Tilde{F}}_W)+ \\ 
   & \nn \quad  +
  B_8\, (-i\, \bm{e}^a\,\bm{\bar{\tilde{\psi}}}\,\Gamma_a\,\bm{\tilde{\rho}}\,\bm{\Tilde{F}}_R + \bm{e}^a\,\bm{\bar{\tilde{\psi}}}\,\Gamma_a\,\Gamma_5\,\bm{\tilde{\rho}}\,\bm{\Tilde{F}}_W)  + 
 B_9 \,(i \,\bm{f}^a\,\bm{e}_a\,\bm{\bar{\tilde{\rho}}}\,\Gamma_5\,\bm{\rho}+ \\ 
 & \nn \quad  - 
   \bm{f}^a\,\bm{\bar{\rho}}\,\Gamma_a\,\Gamma_5\,\bm{\psi}\,\bm{\Tilde{F}}_W)   + 
 B_{10}\, (-i\, \bm{e}^a\,\bm{\bar{\tilde{\psi}}}\,\Gamma_a\,\bm{\tilde{\rho}}\, \bm{\Tilde{F}}_R- 
    i\, \bm{\bar{\tilde{\psi}}}\,\Gamma_5\,\bm{\psi}\,\bm{\Tilde{F}}_R^2  + 
   \bm{\bar{\tilde{\psi}}}\,\bm{\psi}\,\bm{\Tilde{F}}_R\,\bm{\Tilde{F}}_W)+ \\ 
   & \nn \quad  + B_{11}\, (\bm{a}\,\bm{b}\,\bm{\Tilde{F}}_R^2 + \bm{a}\,\bm{b}\,\bm{\Tilde{F}}_W^2) + 
 B_{12}\, (\bm{\bar{\tilde{\psi}}}\,\Gamma_5\,\bm{\psi}\,\bm{\Tilde{F}}_R^2 + 
    \bm{\bar{\tilde{\psi}}}\,\Gamma_5\,\bm{\psi}\,\bm{\Tilde{F}}_W^2) + \\ 
    & \nn  \quad +
 B_{13}\,(2\, \bm{\bar{\tilde{\psi}}}\,\bm{\psi}\,\bm{\bar{\tilde{\rho}}}\,\Gamma_5\,\bm{\rho} - 
    2\, \bm{\bar{\tilde{\psi}}}\,\Gamma_5\,\bm{\psi}\,\bm{\bar{\tilde{\rho}}}\,\bm{\rho}  - \bm{\bar{\tilde{\psi}}}\,\Gamma^a\,\bm{\psi}\,\bm{\bar{\tilde{\rho}}}\,\Gamma_a\,\Gamma_5\,\bm{\rho}
   + \bm{\bar{\tilde{\psi}}}\,\Gamma^a\,\Gamma_5\,\bm{\psi}\,\bm{\bar{\tilde{\rho}}}\,\Gamma_a\,\bm{\rho} + \\ 
   & \nn \quad  + 
    i \,\bm{e}_a\,\bm{\bar{\tilde{\psi}}}\,\Gamma^a\,\bm{\tilde{\rho}}\,\bm{\Tilde{F}}_R + \bm{e}^b\,\bm{\bar{\tilde{\psi}}}\,\Gamma_a\,\Gamma_5\,\bm{\tilde{\rho}}\,\bm{\tilde{R}}^{a}\,_{b})  +
 B_{14}\, (-2\,\bm{f}^a\,\bm{e}_a\,\bm{\bar{\tilde{\rho}}}\,\Gamma_5\,\bm{\rho}   +
   \bm{f}^a\,\bm{e}^b\,\bm{\Tilde{F}}_R\,\bm{\tilde{R}}_{ab})+ \\ 
   & \nn \quad + 
 B_{15}\, (-4\,\bm{f}^a\,\bm{e}_a\,\bm{\bar{\tilde{\rho}}}\,\Gamma_5\,\bm{\rho}  + 
   \epsilon_{abcd}\, \bm{f}^c\bm{e}^d\,\bm{\Tilde{F}}_W\,\bm{\tilde{R}}^{ab}) + \\ 
   & \nn \quad  +
 B_{16}\, (-4 i\, \bm{\bar{\tilde{\psi}}}\,\bm{\psi}\,\bm{\bar{\tilde{\rho}}} \,\Gamma_5\,\bm{\rho} + 
    4 i\, \bm{\bar{\tilde{\psi}}}\,\Gamma_5\,\bm{\psi}\,\bm{\bar{\tilde{\rho}}}\, \bm{\rho}  - 
    2 i \,\bm{\bar{\tilde{\psi}}}\,\Gamma^a\,\bm{\psi}\,\bm{\bar{\tilde{\rho}}} \,\Gamma_a\,\Gamma_5\,\bm{\rho} + \\ 
    & \nn \quad  + 
    2 i\, \bm{\bar{\tilde{\psi}}}\,\Gamma^a\,\Gamma_5\,\bm{\psi}\,\bm{\bar{\tilde{\rho}}}\, \Gamma_a\,\bm{\rho}  +
    2 \,\bm{e}^a\,\bm{\bar{\tilde{\psi}}}\,\Gamma_a\,\bm{\tilde{\rho}}\,\bm{\Tilde{F}}_R  + \epsilon_{abcd}\,\bm{e}^d\,\bm{\bar{\tilde{\psi}}}\,\Gamma^a\,\bm{\tilde{\rho}}\,\bm{\tilde{R}}^{bc})  + \\ 
    & \nn \quad  +
 B_{17}\, (\bm{f}^b\,\bm{\bar{\rho}}\,\Gamma_a\,\Gamma_5\,\bm{\psi}\,\bm{\tilde{R}}^{a}\,_{b} + \frac{1}{2}i\,\epsilon_{abcd}\, \bm{f}^d\,\bm{\bar{\rho}}\,\Gamma^a\,\bm{\psi}\,\bm{\tilde{R}}^{bc}) \\ 
 & \nn \quad  + 
 B_{18}\, (4 \,\bm{f}^a\,\bm{e}_a\,\bm{\bar{\tilde{\rho}}}\,\Gamma_5\,\bm{\rho} -
    2\,\epsilon_{abcd}\,\bm{f}^d\,\bm{\bar{\rho}}\,\Gamma^a\,\bm{\psi}\,\bm{\tilde{R}}^{bc} - \epsilon_{abce}\,\bm{f}^e\bm{e}^d\bm{\tilde{R}}^{ab}\bm{\tilde{R}}^{c}\,_{d})+ \\ 
    & \nn \quad    +
 B_{19}\, \bm{a}\,\bm{\bar{\tilde{\psi}}}\,\Gamma_a\,\bm{\tilde{\rho}}\,\bm{T}^a + 
 B_{20}\, \bm{b}\,\bm{\bar{\tilde{\psi}}}\,\Gamma_a\,\Gamma_5\,\bm{\tilde{\rho}}\,\bm{T}^a  + 
 B_{21} \,\bm{\bar{\tilde{\psi}}}\,\Gamma_a\,\bm{\tilde{\psi}}\,\bm{\Tilde{F}}_R\,\bm{T}^a  + 
 B_{22} \,\bm{f}_a\,\bm{\bar{\tilde{\psi}}}\,\Gamma_5\,\bm{\rho}\,\bm{T}^a + \\ 
 & \nn \quad + B_{23}\, \bm{f}^b\,\bm{\bar{\tilde{\psi}}}\,\Gamma_{ab}\,\Gamma_5\,\bm{\rho}\,\bm{T}^a    +
 B_{24}\,\bm{f}_a\,\bm{\bar{\tilde{\rho}}}\,\Gamma_5\,\bm{\psi}\,\bm{T}^a  + 
 B_{25} \,\bm{f}^b\,\bm{\bar{\tilde{\rho}}}\,\Gamma_{ab}\,\Gamma_5\,\bm{\psi}\,\bm{T}^a  +\\ 
 & \nn \quad  + 
 B_{26}\,\bm{f}_a\,\bm{b}\,\bm{\Tilde{F}}_R\,\bm{T}^a  + 
 B_{27}\,\bm{f}_a\,\bm{a}\,\bm{\Tilde{F}}_W\,\bm{T}^a   + 
 B_{28}\,\bm{f}_b\,\bm{a}\,\bm{\tilde{R}}_{a}\,^{b}\,\bm{T}^a  +
 B_{29} \,\epsilon_{abcd}\,\bm{f}^d\bm{b}\bm{\tilde{R}}^{bc}\bm{T}^a \\ 
 & \nn \quad  + B_{30}\, \epsilon_{abcd}\,\bm{\bar{\tilde{\psi}}}\,\Gamma^a\,\bm{\tilde{\psi}}\,\bm{\tilde{R}}^{cd}\bm{T}^b  + 
 B_{31}\,\bm{a}\,\bm{b}\,\bm{T}^a\,\bm{\tilde{T}}_a + 
 B_{32}\, \bm{\bar{\tilde{\psi}}}\,\Gamma_5\,\bm{\psi}\,\bm{T}^a\bm{\tilde{T}}_a \\ 
 & \nn  \quad   +
 B_{33}\, (-\frac{1}{2} i \,\bm{\bar{\tilde{\psi}}}\,\Gamma^a\,\bm{\psi}\,\bm{\bar{\tilde{\rho}}}\,\Gamma_a\,\Gamma_5\,\bm{\rho} + 
    \frac{1}{2} i \,\bm{\bar{\tilde{\psi}}}\,\Gamma^a\,\Gamma_5\,\bm{\psi}\, \bm{\bar{\tilde{\rho}}}\,\Gamma_a\,\bm{\rho} + 
   \\ 
   & \nn \quad + \frac{1}{8} i\, \bm{\bar{\tilde{\psi}}}\,\Gamma_{ab}\,\bm{\psi}\,\bm{\Tilde{F}}_R\,\bm{\tilde{R}}^{ab}  - 
    \frac{1}{8} \,\bm{\bar{\tilde{\psi}}}\,\Gamma_{ab}\,\Gamma_5\,\bm{\psi}\,\bm{\Tilde{F}}_W\,\bm{\tilde{R}}^{ab} + \bm{e}_a \,\bm{\bar{\tilde{\psi}}}\,\Gamma_5\,\bm{\rho}\,\bm{\tilde{T}}^a) + \\ 
    & \nn \quad  +
 B_{34}\, ( \frac{1}{2} i\, \bm{\bar{\tilde{\psi}}}\,\Gamma^a\,\bm{\psi}\, \bm{\bar{\tilde{\rho}}}\,\Gamma_a\,\Gamma_5\,\bm{\rho} - 
    \frac{1}{2} i \,\bm{\bar{\tilde{\psi}}}\,\Gamma^a\,\Gamma_5\,\bm{\psi}\, \bm{\bar{\tilde{\rho}}}\,\Gamma_a\,\bm{\rho}   + 
   \\ & \nn 
   \quad  - \frac{1}{8} i\, \bm{\bar{\tilde{\psi}}}\,\Gamma_{ab}\,\bm{\psi} \,\bm{\Tilde{F}}_R\,\bm{\tilde{R}}^{ab} + 
    \frac{1}{8}\, \bm{\bar{\tilde{\psi}}}\,\Gamma_{ab}\,\Gamma_5\,\bm{\psi} \,\bm{\Tilde{F}}_W\,\bm{\tilde{R}}^{ab} + 
   \bm{e}^b\, \bm{\bar{\tilde{\psi}}}\,\Gamma_{ab}\,\Gamma_5\,\bm{\rho}\,\bm{\tilde{T}}^a ) + \\ 
   & \nn \quad  + B_{35}\, (\bm{\bar{\psi}}\,\Gamma_a\,\bm{\psi}\,\bm{\Tilde{F}}_R \, \bm{\tilde{T}}^a + 
  \bm{e}_a \,\bm{\bar{\tilde{\rho}}}\,\Gamma_5\, \bm{\psi}\,\bm{\tilde{T}}^a  - 
   \bm{e}^b \,\bm{\bar{\tilde{\rho}}}\,\Gamma_{ab}\,\Gamma_5 \,\bm{\psi}\bm{\tilde{T}}^a) + \\ 
   & \nn \quad  +
 B_{36}\, (i \,\bm{e}^a\,\bm{b}\,\bm{\bar{\tilde{\rho}}}\,\Gamma_a\,\Gamma_5\,\bm{\tilde{\rho}} - 
    i \,\bm{b}\,\bm{\bar{\tilde{\rho}}}\,\bm{\psi}\,\bm{\Tilde{F}}_R + 
   \bm{b}\, \bm{\bar{\tilde{\rho}}}\,\Gamma_5\,\bm{\psi}\,\bm{\Tilde{F}}_W + \bm{e}_a\,\bm{b}\,\bm{\Tilde{F}}_R\,\bm{\tilde{T}}^a) + \\ 
   & \nn \quad  +
 B_{37}\, (\bm{a}\,\bm{\bar{\tilde{\rho}}}\,\Gamma_5\,\bm{\psi}\,\bm{\Tilde{F}}_R+ i \,\bm{a}\,\bm{\bar{\tilde{\rho}}}\,\bm{\psi}\,\bm{\Tilde{F}}_W + 
   \bm{e}_a\,\bm{a}\, \bm{\Tilde{F}}_W\,\bm{\tilde{T}}^a) + \\ 
   & \nn \quad  +
 B_{38}\, (\bm{a}\,\bm{\bar{\tilde{\rho}}}\,\Gamma_5\,\bm{\psi}\,\bm{\Tilde{F}}_R +i \,\bm{a}\,\bm{\bar{\tilde{\rho}}}\,\bm{\psi}\,\bm{\Tilde{F}}_W + 
    2\, \bm{a}\,\bm{\bar{\rho}}\,\Gamma^a\,\bm{\psi}\,\bm{\tilde{T}}_a  + \bm{a}\,\bm{e}_b\bm{\tilde{R}}_{a}\,^{b}\,\bm{\tilde{T}}^a) + \\ 
    & \nn \quad  +
 B_{39}\, (-2 i\,\bm{e}^a\,\bm{b}\,\bm{\bar{\tilde{\rho}}}\,\Gamma_a\,\Gamma_5\, \bm{\tilde{\rho}} + 
    2 i\, \bm{b}\,\bm{\bar{\tilde{\rho}}}\,\bm{\psi}\,\bm{\Tilde{F}}_R - 
    2\, \bm{b}\,\bm{\bar{\tilde{\rho}}}\,\Gamma_5\,\bm{\psi}\,\bm{\Tilde{F}}_W - 4 i\, \bm{b} \,\bm{\bar{\rho}}\,\Gamma^a\,\Gamma_5\,\bm{\psi}\bm{\tilde{T}}_a  + \\ 
    &  \quad   + \epsilon_{abcd}\,
    \bm{e}^d\,\bm{b}\,\bm{\tilde{R}}^{bc}\,\bm{\tilde{T}}^a) +
 B_{40}\,\epsilon_{abcd}\,\bm{f}^c\bm{e}^d\,\bm{T}^a\bm{\tilde{T}}^b  + 
 B_{41}\, \bm{\bar{\tilde{\psi}}}\,\Gamma_{ab}\,\Gamma_5\,\bm{\psi}\,\bm{T}^a\,\bm{\tilde{T}}^b 
\end{align}
\begin{align}\label{secscg:w65}
  & \nn     W_{6,5}=C_1\, (\epsilon_{abcd}\,\bm{f}^b\bm{e}^c\bm{e}^d\,\bm{\bar{\tilde{\psi}}}\,\Gamma^a\,\bm{\tilde{\rho}}  - 
    2 i\,  \bm{f}^b\bm{e}^a\bm{e}_b\,\bm{\bar{\tilde{\psi}}}\,\Gamma_a\,\Gamma_5\, \bm{\tilde{\rho}} - 
    i\, \bm{f}^a\bm{e}_a\,\bm{\bar{\tilde{\psi}}}\,\bm{\psi}\,\bm{\Tilde{F}}_R  + \\ 
    & \nn \quad - i\,  \bm{f}^a\bm{e}^b \,\bm{\bar{\tilde{\psi}}}\,\Gamma_{ab}\,\bm{\psi}\,\bm{\Tilde{F}}_R + 
   \bm{f}^a \bm{e}_a \,\bm{\bar{\tilde{\psi}}}\,\Gamma_5\,\bm{\psi}\,\bm{\Tilde{F}}_W + 
   \bm{f}^a\bm{e}^b \, \bm{\bar{\tilde{\psi}}}\,\Gamma_{ab}\,\Gamma_5\,\bm{\psi}\,\bm{\Tilde{F}}_W) + \\ 
   & \nn \quad  +C_2\, (\bm{f}^a\,\bm{a}\,\bm{e}_a\,\bm{\bar{\tilde{\psi}}}\,\bm{\rho} +
    \bm{f}^a\,\bm{a}\,\bm{e}^b\,\bm{\bar{\tilde{\psi}}}\,\Gamma_{ab}\,\bm{\rho} )  + 
 C_3\, (\bm{f}^a\bm{e}_a\,\bm{b}\,\bm{\bar{\tilde{\psi}}}\,\Gamma_5\,\bm{\rho}\, +\bm{f}^a\bm{e}^b\,\bm{b}\,\bm{\bar{\tilde{\psi}}}\,\Gamma_{ab}\,\Gamma_5\,\bm{\rho}) + \\ 
 & \nn \quad  + C_{4}\, (-\epsilon_{abcd}\,\bm{f}^b\bm{f}^c\bm{e}^d\,\bm{\bar{\rho}}\,\Gamma^a\,\bm{\psi} + 
    2 i\,\bm{f}^a\bm{f}^b\bm{e}_b\,\bm{\bar{\rho}}\,\Gamma_a\,\Gamma_5\,\bm{\psi}) + \\ 
    & \nn \quad  +
 C_{5} \,(-2\, \bm{f}^a\bm{f}^b\bm{e}_a\bm{e}_b\,\bm{\Tilde{F}}_R + \epsilon^{abcd}\,\bm{f}^a\bm{f}^b\bm{e}^c\bm{e}^d\,\bm{\Tilde{F}}_W) + \\ 
 & \nn \quad +
 C_{6}\, (2 \,\bm{a}\,\bm{\bar{\tilde{\psi}}}\,\Gamma^a\,\bm{\tilde{\psi}}\,\bm{\bar{\rho}}\,\Gamma_a\,\bm{\psi} + \bm{a}\,\bm{e}^a\,\bm{\bar{\tilde{\psi}}}\,\Gamma_a\,\bm{\tilde{\psi}}\,\bm{\Tilde{F}}_W + \bm{a}\,\bm{e}_b\,\bm{\bar{\tilde{\psi}}}\,\Gamma_a\,\bm{\tilde{\psi}}\,\bm{\tilde{R}}^{ab}) + \\ 
 & \nn \quad  +
 C_{7}\, (\frac{3}{2}\, \bm{f}^a\,\bm{\bar{\tilde{\psi}}}\,\Gamma_5\,\bm{\rho}\,\bm{\bar{\psi}}\,\Gamma_a\, \bm{\psi} - \frac{1}{2}\,\bm{f}^b\,\bm{\bar{\tilde{\psi}}}\,\Gamma_{ab}\,\Gamma_5\,\bm{\rho}\,\bm{\bar{\psi}}\,\Gamma^a\,\bm{\psi} +\frac{1}{2}\,\bm{f}^b\,\bm{\bar{\tilde{\psi}}}\,\Gamma^a\,\Gamma_5\,\bm{\rho}\,\bm{\bar{\psi}}\,\Gamma_{ab}\,\bm{\psi} +\\ 
 & \nn \quad  - \frac{1}{2}\,\bm{f}^b\,\bm{\bar{\tilde{\psi}}}\,\Gamma^a\,\bm{\rho}\,\bm{\bar{\psi}}\,\Gamma_{ab}\,\Gamma_5\,\bm{\psi}  - \bm{f}^a\bm{e}^b\,\bm{\bar{\tilde{\psi}}}\,\Gamma_{ab}\,\Gamma_5\,\bm{\psi}\,\bm{\Tilde{F}}_W + \bm{f}^b\bm{e}_c\,\bm{\bar{\tilde{\psi}}}\,\Gamma_{ab}\,\Gamma_5\,\bm{\psi}\,\bm{\tilde{R}}^{ac}) + \\ 
 & \nn \quad  + C_8\, (-\frac{3}{2} \,\bm{f}^a\,\bm{\bar{\tilde{\psi}}}\,\Gamma_5\,\bm{\rho}\,\bm{\bar{\psi}}\,\Gamma^a\, \bm{\psi} + \frac{1}{2}\,\bm{f}^b\,\bm{\bar{\tilde{\psi}}}\,\Gamma_{ab}\,\Gamma_5\,\bm{\rho}\,\bm{\bar{\psi}}\,\Gamma^a\,\bm{\psi} + \frac{1}{2}\,\bm{f}^b\,\bm{\bar{\tilde{\psi}}}\,\Gamma^a\,\Gamma_5\,\bm{\rho}\,\bm{\bar{\psi}}\,\Gamma_{ab}\,\bm{\psi}+ \\ 
 & \nn \quad  - \frac{1}{2}\,\bm{f}^b\,\bm{\bar{\tilde{\psi}}}\,\Gamma^a\,\bm{\rho}\,\bm{\bar{\psi}}\,\Gamma_{ab}\,\Gamma_5\,\bm{\psi} 
   + i\,\bm{f}^a\bm{e}^b\,\bm{\bar{\tilde{\psi}}}\,\Gamma_{ab}\,\bm{\psi}\,\bm{\Tilde{F}}_R
     - \frac{1}{2}\,\bm{f}^c\bm{e}_c\,\bm{\bar{\tilde{\psi}}}\,\Gamma_{ab}\,\Gamma_5\,\bm{\psi}\,\bm{\tilde{R}}^{ab} + \\ 
     & \nn \quad + \bm{f}_c\bm{e}^b\,\bm{\bar{\tilde{\psi}}}\,\Gamma_{ab}\,\Gamma_5\,\bm{\psi}\,\bm{\tilde{R}}^{ac})  +
 C_9\, (-2\, \bm{f}^a\,\bm{a}\,\bm{b}\,\bm{\bar{\rho}}\, \Gamma_a\,\bm{\psi} + 
   \bm{f}^a\,\bm{a}\,\bm{e}_a\,\bm{b} \,\bm{\Tilde{F}}_W  + \bm{f}_a\,\bm{a}\,\bm{e}_b\,\bm{b}\,\bm{\tilde{R}}^{ab}) +\\ 
   & \nn  \quad  +
 C_{10}\, (-4 i \,\bm{f}^a\bm{f}^b\bm{e}_b\, \bm{\bar{\rho}}\,\Gamma_a\,\Gamma_5\,\bm{\psi} + 
    2 \,\bm{f}^a \bm{f}^b \bm{e}_a\bm{e}_b \,\bm{\Tilde{F}}_R +\epsilon_{acde}\, \bm{f}^c\bm{f}^d\bm{e}^b\bm{e}^e \bm{\tilde{R}}^{a}\,_{b}) +\\ 
    & \nn  \quad  +
  C_{11} \,(-i\, \bm{f}^a\,\bm{\bar{\tilde{\psi}}}\,\Gamma_5\,\bm{\rho}\,\bm{\bar{\psi}} \,\Gamma_a\,\bm{\psi} - 
    i \,\bm{f}^b\,\bm{\bar{\tilde{\psi}}}\,\Gamma_{ab}\,\Gamma_5\,\bm{\rho}\,\bm{\bar{\psi}}\, \Gamma^a\,\bm{\psi} + 
    i \,\bm{f}^b\,\bm{\bar{\tilde{\psi}}}\,\Gamma^{a}\,\Gamma_5\,\bm{\rho}\,\bm{\bar{\psi}}\, \Gamma_{ab}\bm{\psi}+ \\ & \nn  \quad - 
    i \,\bm{f}^b\,\bm{\bar{\tilde{\psi}}}\,\Gamma^{a}\,\bm{\rho}\,\bm{\bar{\psi}}\, \Gamma_{ab}\,\Gamma_5\,\bm{\psi}  +
    2 \, \bm{f}^a\bm{e}_a\,\bm{\bar{\tilde{\psi}}}\,\bm{\psi} \,\bm{\Tilde{F}}_R+\epsilon_{abcd}\,\bm{f}^c\bm{e}^d \,\bm{\bar{\tilde{\psi}}}\,\bm{\psi}\,\bm{\tilde{R}}^{ab}) + \\ 
    & \nn \quad + C_{12}\,(\frac{1}{2}\,\bm{f}^a\,\bm{\bar{\tilde{\psi}}}\,\Gamma_5\,\bm{\rho}\,\bm{\bar{\psi}}\, \Gamma_a\,\bm{\psi}  + \frac{1}{2}\,\bm{f}^b\,\bm{\bar{\tilde{\psi}}}\,\Gamma_{ab}\,\Gamma_5\,\bm{\rho}\,\bm{\bar{\psi}}\, \Gamma^a\,\bm{\psi}
     + \frac{1}{2}\,\bm{f}^b\,\bm{\bar{\tilde{\psi}}}\,\Gamma^{a}\,\Gamma_5\,\bm{\rho}\,\bm{\bar{\psi}}\, \Gamma_{ab}\,\bm{\psi}
   + \\ 
   & \nn  \quad    - \frac{1}{2}\,\bm{f}^b\,\bm{\bar{\tilde{\psi}}}\,\Gamma^{a}\,\bm{\rho}\,\bm{\bar{\psi}}\, \Gamma_{ab}\,\Gamma_5\,\bm{\psi}
     +\bm{f}^a\bm{e}_a\,\bm{\bar{\tilde{\psi}}}\,\Gamma_5\,\bm{\psi}\, \bm{\Tilde{F}}_W + \bm{f}_a\bm{e}_b\, \bm{\bar{\tilde{\psi}}}\,\Gamma_5\,\bm{\psi}\,\bm{\tilde{R}}^{ab}) + \\ 
     & \nn \quad  +
 C_{13}\, (-4 i\, \bm{b}\, \bm{\bar{\tilde{\psi}}}\,\Gamma^a\,\bm{\tilde{\psi}}\,\bm{\bar{\rho}}\, \Gamma_a\,\Gamma_5\,\bm{\psi} + 
    2 \,\bm{e}^a\,\bm{b}\,\bm{\bar{\tilde{\psi}}}\,\Gamma_a\,\bm{\tilde{\psi}}\, \bm{\Tilde{F}}_R+ \epsilon_{abcd}\,\bm{e}^d\,\bm{b}\,\bm{\bar{\tilde{\psi}}}\,\Gamma^a\,\bm{\tilde{\psi}}\, \bm{\tilde{R}}^{bc})+ \\ 
    & \nn \quad   + 
 C_{14}\, \bm{a}\,\bm{b}\,\bm{\bar{\tilde{\psi}}}\,\Gamma_a\,\bm{\tilde{\psi}}\,\bm{T}^a  +
 C_{15}\,\bm{f}_a\bm{f}^b\,\bm{a}\,\bm{e}_b\,\bm{T}^a + C_{16}\,\epsilon_{abcd}\,\bm{f}^b\bm{f}^c\bm{e}^d\,\bm{b}\,\bm{T}^a +  \\ &\nn  \quad +
 C_{17}\,  \bm{f}_a\,\bm{a}\,\bm{\bar{\tilde{\psi}}}\,\bm{\psi}\,\bm{T}^a + 
 C_{18}\,\bm{f}^b\,\bm{a}\, \bm{\bar{\tilde{\psi}}}\,\Gamma_{ab}\,\bm{\psi}\,\bm{T}^a  + 
 C_{19}\, \bm{f}_a\,\bm{b}\, \bm{\bar{\tilde{\psi}}}\,\Gamma_5\,\bm{\psi}\,\bm{T}^a + \\ & \nn \quad  +
 C_{20}\, \bm{f}^b\,\bm{b}\, \bm{\bar{\tilde{\psi}}}\,\Gamma_{ab}\,\Gamma_5\,\bm{\psi}\,\bm{T}^a + 
 C_{21} \,\epsilon_{abcd}\,\bm{f}^c\bm{e}^d \,\bm{\bar{\tilde{\psi}}}\,\Gamma^a\,\bm{\tilde{\psi}}\,\bm{T}^b + 
 C_{22} \,\epsilon_{abcd}\,\bm{f}^c\bm{f}^d\,\bm{\bar{\psi}}\,\Gamma^a\,\bm{\psi}\,\bm{T}^b  + \\ & \nn  \quad +
 C_{23}\, (-2\, \bm{f}^a\bm{e}_a\,\bm{b}\,\bm{\bar{\tilde{\rho}}}\,\Gamma_5\,\bm{\psi}  +\bm{f}^a\bm{e}^b\,\bm{b}\,\bm{\bar{\tilde{\rho}}}\,\Gamma_{ab}\,\Gamma_5\,\bm{\psi} + 2\,\bm{f}^a\,\bm{b}\,\bm{\bar{\psi}}\,\Gamma_a\,\bm{\psi}\,\bm{\Tilde{F}}_R
    + \\ 
    & \nn  \quad  +\epsilon_{abcd}\,\bm{f}^b\bm{e}^c\bm{e}^d\,\bm{b}\,\bm{\tilde{T}}^a)  +
 C_{24}\, (-2\, \bm{e}^a\,\bm{\bar{\tilde{\psi}}}\,\Gamma_a\, \bm{\tilde{\psi}}\, \bm{\bar{\tilde{\rho}}}\,\Gamma_5\,\bm{\psi} +2\, \bm{e}^b  \,\bm{\bar{\tilde{\psi}}}\,\Gamma^a\, \bm{\tilde{\psi}}\, \bm{\bar{\tilde{\rho}}}\,\Gamma_{ab}\,\Gamma_5\,\bm{\psi}
      + \\ 
 & \quad   -  2 \,\bm{\bar{\tilde{\psi}}}\,\Gamma_a\,\bm{\tilde{\psi}}\,\bm{\bar{\psi}}\, \Gamma^a \,\bm{\psi}\,\bm{\Tilde{F}}_R +\epsilon_{abcd}\, \bm{e}^c\bm{e}^d\,\bm{\bar{\tilde{\psi}}}\,\Gamma^a\,\bm{\tilde{\psi}}\, \bm{\tilde{T}}^b) + C_{25}\,(\bm{\bar{\tilde{\psi}}}\,\Gamma_a\, \bm{\tilde{\psi}}\,\bm{\bar{\tilde{\psi}}}\,\Gamma_5 \,\bm{\psi} \,\bm{T}^a)
\end{align}
\begin{align}\label{secscg:w53}
        W_{5,3}& \nn =a_1 \,  \epsilon_{abcd}\,\bm{f}^a\bm{\tilde{R}}^{cd}\bm{T}^b  +  a_2\, (i \,\bm{e}^a \bm{\bar{\tilde{\rho}}}\,\Gamma_{a}\,\Gamma_5\,\bm{\tilde{\rho}} + 
    i\, \bm{\bar{\tilde{\rho}}}\,\bm{\psi}\,\bm{\Tilde{F}}_R - 
   \bm{\bar{\tilde{\rho}}}\,\Gamma_5\, \bm{\psi}\,\bm{\Tilde{F}}_W +\bm{e}^a\,\bm{\Tilde{F}}_R\,\bm{\tilde{T}}_a)+ \\ & \nn  +
 a_3 \,\bm{a}\,\bm{T}^a \bm{\tilde{T}}_a 
+ a_4\, (2 i \,\bm{e}^a\,\bm{\bar{\tilde{\rho}}}\,\Gamma_a\,\Gamma_5\,\bm{\tilde{\rho}}  + 
    2 i\, \bm{\bar{\tilde{\rho}}}\,\bm{\psi}\,\bm{\Tilde{F}}_R - 
    2\, \bm{\bar{\tilde{\rho}}}\,\Gamma_5\,\bm{\psi}\, \bm{\Tilde{F}}_W+ \\ & \nn -  4 i\, \bm{\bar{\rho}}\,\Gamma_a\,\Gamma_5\,\bm{\psi}\,\bm{\tilde{T}}^a  + 
    \epsilon_{abcd}\,\bm{e}^a\bm{\tilde{R}}^{cd}\bm{\tilde{T}}^b) +
    a_5\, \bm{f}^a\,\bm{\bar{\rho}}\,\Gamma_a\,\Gamma_5\, \bm{\rho} + a_6\, (\bm{a}\,\bm{\Tilde{F}}_R^2 + \bm{a}\,\bm{\Tilde{F}}_W^2) + \\ & + 
 a_7\, (-i\, \bm{\bar{\tilde{\psi}}}\,\bm{\rho}\,\bm{\Tilde{F}}_R+ 
   \bm{\bar{\tilde{\psi}}}\,\Gamma_5\,\bm{\rho}\,\bm{\Tilde{F}}_W) + 
 a_8 \,\bm{f}^a \,\bm{\Tilde{F}}_R\, \bm{T}_a + 
 a_9 \, \bm{\bar{\tilde{\psi}}}\,\Gamma^a\,\Gamma_5\,\bm{\tilde{\rho}}\,\bm{T}_a 
\end{align}
\begin{align}\label{secscg:w54}
        W_{5,4}& \nn  =b_1 \,  (2\, \bm{f}^a\,\bm{a}\,\bm{\bar{\rho}}\,\Gamma_a\,\bm{\psi} + 
   \bm{f}^a\,\bm{a}\,\bm{e}_a\,\bm{\Tilde{F}}_W + 
    \bm{f}_a\,\bm{a}\,\bm{e}_b\,\bm{\tilde{R}}^{ab}) +\\ & \nn +
      b_2 \,(2\, \bm{f}^a\bm{e}_a\,\bm{\bar{\tilde{\rho}}}\,\Gamma_5\,\bm{\psi} - 
    2\, \bm{f}^a\,\bm{e}^b\,\bm{\bar{\tilde{\rho}}}\,\Gamma_{ab}\,\Gamma_5\,\bm{\psi} + 
    2 \, \bm{f}^a\,\bm{\bar{\psi}}\,\Gamma_a\,\bm{\psi}\,\bm{\Tilde{F}}_R +\epsilon_{abcd} \,\bm{f}^b\,\bm{e}^c\,\bm{e}^d\,\bm{\tilde{T}}^a)+ \\ & \nn + 
 b_3\, (-4 \,i\,\bm{f}^a\,\bm{b}\,\bm{\bar{\rho}}\,\Gamma^a\,\Gamma_5\,\bm{\psi} + 
    2\, \bm{f}^a\bm{e}_a\,\bm{b}\,\bm{\Tilde{F}}_R+ \epsilon_{abcd}
    \,\bm{f}^c\bm{e}^d\,\bm{b}\,\bm{\tilde{R}}^{ab}) + \\ & \nn +
 b_4 \,(4\, i\, \bm{\bar{\tilde{\psi}}}\,\Gamma^a\,\bm{\tilde{\psi}}\,\bm{\bar{\rho}}\,\Gamma_a\,\Gamma_5\,\bm{\psi}  + 
    2\, \bm{e}^a\,\bm{\bar{\tilde{\psi}}}\,\Gamma_a\,\bm{\tilde{\psi}}\,\bm{\Tilde{F}}_R  +\epsilon_{abcd}\, \bm{e}^b\,\bm{\bar{\tilde{\psi}}}\, \Gamma^a\,\bm{\psi}\,\bm{\tilde{R}}^{cd}) + \\ & \nn +
 b_5\, \bm{f}^a\,\bm{\bar{\tilde{\psi}}}\,\Gamma_5\,\bm{\psi}\bm{T}_a  + 
 b_6 \,\bm{a}\,\bm{\bar{\tilde{\psi}}}\,\Gamma_a\,\bm{\tilde{\psi}}\,\bm{T}^a  +\\ & \nn + 
 b_7 \,\bm{f}_a\,\bm{a}\,\bm{b}\,\bm{T}^a + 
 b_8 \,\epsilon_{abcd}\,\bm{f}^b\,\bm{f}^c\,\bm{e}^d\,\bm{T}^a + 
 b_{9} \, \bm{f}^a\,\bm{\bar{\tilde{\psi}}}\,\Gamma_{ab}\,\Gamma_5\,\bm{\psi}\,\bm{T}^b + \\ & +
b_{10}\, (\bm{f}^a\bm{e}_a\,\bm{\bar{\tilde{\psi}}}\,\Gamma_5\,\bm{\rho} + 
  \bm{f}^a\bm{e}^b\, \bm{\bar{\tilde{\psi}}}\,\Gamma_{ab}\,\Gamma_5\,\bm{\rho})
\end{align}

\subsection{Cohomological Analysis}
By inserting Eqs. (\ref{secscg:w63}-\ref{secscg:w54}) into the descent equations (\ref{seccg:kerneldeltadescent}) we computed the closed  elements in $\ker_\delta(W_6^{\text{super-conf}})$: we exhibit them explicitly in Appendix \ref{appendix:superconformalcocycle}.  We find 
  \begin{equation}
  \dim \ker_\delta(W^{\text{super-conf}}_6) = 18
  \end{equation}
  One can  similarly verify that 
    \begin{equation}
  \dim \ker_\delta(W^{\text{super-conf}}_5) = 3 \Rightarrow \dim \Im_\delta(W^{\text{super-conf}}_5) = 16
  \end{equation}
Hence the  cohomology $H_\delta(W^{\text{super-conf}}_6)$ has dimension 2,  just as in the bosonic case
  \begin{equation}
    \dim H_\delta(W^{\text{super-conf}}_6)=   \dim \ker_\delta(W^{\text{super-conf}}_6)-\dim \Im_\delta(W^{\text{super-conf}}_5) = 2
  \end{equation}
  
  \subsection{Anomaly Representatives}\label{sec:scganomalyreprs}

$H_\delta(W_6)$ admits  basis made of homogeneous representatives which are characterized analogously to the bosonic case.  

One representative  is Chern-like, it is a cubic polynomials of the curvatures  with  $W_{6,5} =W_{6,4}=0$
\begin{equation}
	\begin{split}
		W_6^{(A)}= W_{6,3}^{(A)} = & \, \frac{1}{\pi^2}\,\bigg(\frac{5}{4}\, (\bm{\Tilde{{F}}_R})^3+\frac{1}{4}\,(\bm{\Tilde{F}_W})^2\bm{\tilde{F}}_R-\frac{1}{16}\,\epsilon_{abcd}\,\bm{\Tilde{R}}^{ab}\,\bm{\Tilde{R}}^{cd}\,\bm{\Tilde{F}_W}+ \\ & -\frac{1}{8}\,\bm{\Tilde{F}_R}\,\tr \bm{\Tilde{R}}^2  -\frac{1}{2}\,\bm{\Bar{\rho}}\,\Gamma_{ab}\,\Gamma_{_5}\, \bm{\Tilde{\rho}}\,\bm{\Tilde{R}}^{ab}+5\,i\,\bm{\Bar{\rho}}\,\bm{\Tilde{\rho}}\,
		\bm{\tilde{F}}_R+\,\bm{\Bar{\rho}}\,\Gamma_{5}\,\bm{\Tilde{\rho}}\,\bm{\Tilde{F}_W}+ \\ & +i\,\bm{\Bar{\rho}}\,\Gamma_a \,\Gamma_{5}\, \bm{\rho}\,\bm{\Tilde{T}^a} 
		+\, \bm{\Tilde{F}}_R\,\bm{T}_a\,\bm{\Tilde{T}}^a-\frac{1}{2}\,\epsilon_{abcd}\,\bm{\Tilde{R}}^{cd}\,\bm{T}^a\,\bm{\Tilde{T}}^b-i\, \bm{\bar{\tilde{\rho}}}\,\Gamma^{a}\,\Gamma_5\,\bm{\tilde{\rho}}\,\bm{T}_a  \bigg).
	\end{split}
\end{equation}
$W^{(A)}_{6,3}$  is  an  \textbf{invariant polynomial}  of the curvatures:  it is  the \textbf{unique} invariant Chern polynomial of  the   $\mathfrak{su}(2,2|1)$ super Lie algebra  found in \cite{Imbimbo:2023sph}.  The so-called $a$ superconformal anomaly  $H_{6}^{(a)}$ is described precisely by this class
\begin{equation}
H_{6}^{(a)} = a \, W^{(A)}_{6,3}
\end{equation}
The second class in $H_\delta(W_6)$ cannot be represented by a Chern curvature polynomial. It  admits a homogenous --- with 2 curvatures and 2 connections ---  representative with $W_{6,5} =W_{6,3}=0$:
\begin{equation}
	\begin{split}
		W_{6}^{(B)}=W_{6,4}^{(B)}= \,\frac{1}{\pi^2}\,\bigl( & -4 \,\bm{\bar{\tilde{\rho}}} \,\Gamma_{a b}  \,
		\Gamma_5 \, \bm{\rho} \, \bm{f}^{a}  \bm{e}^{b} + 
		2\, i\, \bm{\bar{\rho}} \,\Gamma_{a}\,  \Gamma_5 \,
		\bm{\rho}\, \bm{f}^{a}\, \bm{b} -  2\, i \,\bm{\bar{\tilde{\rho}}} \,  \Gamma_{a} \,  \Gamma_5 \,
		\bm{\Tilde{\rho}}\, \bm{e}^{a}\, \bm{b}+ \\ &  - 2\, i\,  \bm{\bar{\Tilde{\psi}}}\,   \bm{\rho}\, \bm{b}\,\bm{\Tilde{F}}_R + 
		2 \, \bm{\bar{\Tilde{\psi}}}\, \Gamma_{a}\,  \bm{\Tilde{\rho}} \,\bm{e}^{a}  \,
		\bm{\Tilde{F}}_R     - 2 \, \bm{\bar{\Tilde{\psi}}}\, \Gamma_5\, \bm{\rho} \,  \bm{a} \,
		\bm{\Tilde{F}}_R+ \\  & - 2\, i \,\bm{\bar{\tilde{\rho}}}\,  \bm{\psi}\,  \bm{b}\, \bm{\Tilde{F}}_R - 
		4\, \bm{\bar{\rho}}  \, \Gamma_{a} \,  \bm{\psi} \,\bm{f}^{a}
		\,\bm{\Tilde{F}}_R - 2\, \bm{\bar{\tilde{\rho}}}\,  \Gamma_5 \,    \bm{\psi} \,\bm{a}\, \bm{\Tilde{F}}_R+ \\  &  
		-2 \, \bm{\bar{\Tilde{\psi}}}\,  \Gamma_5 \, \bm{\psi}\,  \bm{\Tilde{F}}_R^2 
		- 2\, i\,  \bm{\bar{\Tilde{\psi}}} \, \bm{\rho}\, \bm{a}\, \bm{\tilde{F}_W}+ 
		2 \, \bm{\bar{\Tilde{\psi}}}\, \Gamma_5\,  \bm{\rho} \,  \bm{b}\,\bm{\tilde{F}_W}+ \\  & - 
		2\, i \, \bm{\bar{\Tilde{\psi}}}\,  \Gamma_{a}  \, \Gamma_5 \, \bm{\Tilde{\rho}} \, \bm{e}^{a}\, \bm{\tilde{F}}_W- 
		2\, i\, \bm{\bar{\tilde{\rho}}}\, \bm{\psi} \,\bm{a}\, \bm{\tilde{F}}_W+ 
		2\, \bm{\bar{\tilde{\rho}}}\,  \Gamma_5 \,  \bm{\psi}  \, \bm{b}\, \bm{\tilde{F}_W}+ \\  & - 2 \,i\, \bm{\bar{\rho}}\,  \Gamma_{a}\,   
		\Gamma_5\,  \bm{\psi} \, \bm{f}^{a}\, \bm{\tilde{F}}_W+  
		\, \bm{f}^{a}\, \bm{e}_{a}\, \bm{\Tilde{F}}_R\, \bm{\tilde{F}_W}- 
		4\, i \, \bm{\bar{\Tilde{\psi}}}\,   \bm{\psi}\, \bm{\Tilde{F}}_R\, \bm{\tilde{F}}_W+ \\  & + 
		2\,  \bm{\bar{\Tilde{\psi}}} \,  \Gamma_5 \,  \bm{\psi}\,  \bm{\tilde{F}_W}^2 +
		\frac{1}{2}\,\epsilon_{abcd} \,\bm{f}^{c} \bm{e}^{d}\, \bm{\tilde{F}_W}\,  \bm{\Tilde{R}}^{a b}   + 2\, i\,  \bm{\bar{\Tilde{\psi}}}\, \Gamma_{a} \,  \Gamma_5 \,\bm{\Tilde{\rho}}\, \bm{b}\, \bm{T}^{a} + \\  &
		+   \,\bm{\bar{\Tilde{\psi}}} \, \Gamma_{a}\,   \bm{\Tilde{\psi}} \,\bm{\Tilde{F}}_R\,  \bm{T}^{a} + 2\,  \bm{\bar{\Tilde{\psi}}} \, \Gamma_{ab}
		\,\Gamma_5 \, \bm{\rho}\,  \bm{f}^{b} \,\bm{T}^{a} - 
		2\, \bm{\bar{\tilde{\rho}}} \,  \Gamma_{a b} \,   
		\Gamma_5  \, \bm{\psi}  \,\bm{f}^{b} \bm{T}^{a} + \\  &+
		\, \bm{f}_{a} \,\bm{b} \,\bm{\Tilde{F}}_R  \,\bm{T}^{a}    - 
		2 \,\bm{f}_{a} \,\bm{a}\, \bm{\tilde{F}}_W \,\bm{T}^{a}  +
		\frac{1}{2}\,\epsilon_{abcd}\, \bm{f}^{d} \,\bm{b} \, \bm{\Tilde{R}}^{b c} \,  \bm{T}^{a}+ \\  & + 2\, i\, \bm{\bar{\rho}}\,  \Gamma_{a}\,   
		\Gamma_5\,   \bm{\psi} \,\bm{b}\,  \bm{\tilde{T}}^{a}
		- 2\,  \bm{\bar{\psi}}\,   \Gamma_{a}\,   \bm{\psi} \, \bm{\Tilde{F}}_R\,  \bm{\tilde{T}}^{a} + 
		2\,  \bm{\bar{\Tilde{\psi}}} \, \Gamma_5 \, \bm{\psi} \, \bm{T}^{a} \,  \bm{\tilde{T}}_{a}+ \\  & +
		2 \, \bm{\bar{\Tilde{\psi}}} \,\Gamma_5 \,  \bm{\rho} \,   \bm{e}_{a} \,
		\bm{\tilde{T}}^{a}    +  2 \, \bm{\bar{\Tilde{\psi}}}\,  \Gamma_{a b}  \,
		\Gamma_5\, \bm{\rho}\,    \bm{e}^{b}\,  \bm{\tilde{T}}^{a}  +
		2 \,\bm{\bar{\tilde{\rho}}} \, \Gamma_5 \,  \bm{\psi}\,
		\bm{e}_{a} \,  \bm{\tilde{T}}^{a} + \\  &-  2\, \bm{\bar{\tilde{\rho}}}\,   \Gamma_{a b} \,  
		\Gamma_5\,   \bm{\psi}\, \bm{e}^{b}\, \bm{\tilde{T}}^{a}   - 
		\, \bm{e}_{a}\, \bm{b}\, \bm{\Tilde{F}}_R\,  \bm{\tilde{T}}^{a}  - 
		2\, \bm{a}\,\bm{e}_{a}\, \bm{\tilde{F}}_W\, \bm{\tilde{T}}^{a} + \\  & +\frac{1}{2}\,  \epsilon_{abcd} \, \bm{e}^{d} \,\bm{b}\,  \bm{\Tilde{R}}^{bc} \,  
		\bm{\tilde{T}}^{a} +  2 \,  \epsilon_{abcd}\, \bm{f}^{c}\,\bm{e}^{d} \, \bm{T}^a  \, 
		\bm{\tilde{T}}^{b}
		\bigr).
	\end{split}
\end{equation}
Just like in the bosonic case the superconformal  $c$-anomaly $H_6^{(c)}$  is described by  a linear combination of the classes  represented by  $ W^{(A)}_{6,3}$  and $W^{(B)}_{6,4}$
\begin{equation}\label{secscg:hcclass}
H_6^{(c)} =  c\, ( W^{(B)}_{6,4}-W^{(A)}_{6,3})
\end{equation}
Therefore the  degree 6 picture of the general  $(a+c)$-superconformal anomaly writes
\begin{equation}\label{seccg:h6ac}
H_6^{(a)}+ H_6^{(c)} = a\,W^{(A)}_{6,3}+  c\, ( W^{(B)}_{6,4}-W^{(A)}_{6,3})  =  (a-c) \,W^{(A)}_{6,3}+ c\, W^{(B)}_{6,4}
\end{equation}
In particular $W^{(B)}_{6,4}$  describes the anomaly  of superconformal theories with $a=c$:
\begin{equation}
H_{6,4}^{(B)}=c\, W^{(B)}_{6,4} = (H_{6}^{(a)} + H_{6}^{(c)})\big|_{a=c}  
\end{equation}
Superconformal theories with $a=c$ are precisely  those which emerge in $AdS_5$ holography: we do not know if the fact that the anomalies of those theories are described by the homogenous class $W^{(B)}_{6,4}$ has an ``a priori'' explanation: we leave this to future investigations. 

Let us turn to the degree 5 picture.  For the  $a$-anomaly 
\begin{equation}
	a\,  W^{(A)}_{6,3} = \delta\, Q^{(a)}_5
\end{equation}
we obtain the unique (up to $\delta$-exact terms) Chern-Simons polynomial of the $\mathfrak{su}(2,2|1)$ super Lie algebra. An explicit representative  $Q^{(a)}_5$ characterized by  renormalization prescriptions that we will specified in the next subsection is
 \begin{subequations}\label{secscg:Q5aanomaly}
	\begin{align}
		Q_{5,3}^{(a)}=& \nn \frac{a}{\pi^2}\,\bigg(\bm{a}\,\bm{\tilde{F}_R}^2
		- \frac{1}{8}\, \bm{a}\,\bm{\tilde{R}}^{ab}\,\bm{\tilde{R}}_{ab}  - 
		\frac{1}{16}\, \epsilon_{abcd} \,\bm{b}\,\bm{\tilde{R}}^{ab}\,\bm{\tilde{R}}^{cd} 
		+ 6\, i\, \bm{\bar{\tilde{\psi}}}\,\bm{\rho}\,\bm{\tilde{F}}_R +\\ &   +
		\frac{1}{2} \, \bm{\bar{\tilde{\psi}}} \,\Gamma_{ab}\,\Gamma_5\, \bm{\rho} \,\bm{\tilde{R}}^{ab}\bigg) \\  
		Q_{5,4}^{(a)}=& \nn \frac{a}{\pi^2}\,\bigg(4\,\bm{\bar{\tilde{\psi}}}\,\Gamma_5\, \bm{\rho}\,\bm{f}^a\,\bm{e}_a+2\,\bm{\bar{\rho}}\,\Gamma_a\, \bm{\psi}\, \bm{f}^a \,\bm{a} + 
		2\,i\,\,  \bm{\bar{\rho}}\,\Gamma_{a} \,\Gamma_5\, \bm{\psi }\,\bm{f}^a\,  \bm{b}  + \\ & \nn+\frac{1}{2}\,\epsilon_{abcd}\,\bm{\bar{\tilde{\psi}}}\,\Gamma^{ab}\,\Gamma_5\,\bm{\tilde{\psi}}\,\bm{\bar{\rho}}\,\Gamma^{cd}\,\Gamma_5\,\bm{\psi}-\bm{f}^a\,\bm{e}_a\,\bm{b}\,\bm{\tilde{F}_R}-
		5 \,  \bm{\bar{\tilde{\psi}}}\,\Gamma_5 \,\bm{\psi}  \,  \bm{a}\,
		\bm{\tilde{F}}_R + \\ &  +
		2 \, \bm{\bar{\tilde{\psi}}}\, \Gamma_a\, \bm{\tilde{\psi}}\, \bm{e}^a\,  \bm{\tilde{F}}_R  -i\, \bm{\bar{\tilde{\psi}}}\,\bm{\psi}\,\bm{a}\,\bm{\tilde{F}_W}- 
		\frac{i }{2} \, \bm{\bar{\tilde{\psi}}} \,\Gamma_{ab} \,\bm{\psi}  \,\bm{a}\, \bm{\tilde{R}}^{ab} -\frac{1}{2}\,\bm{\bar{\tilde{\psi}}}\,\Gamma_{ab}\,\Gamma_5\, \bm{\psi} \,\bm{b}\,\bm{\tilde{R}}^{ab}\bigg) \\
		Q_{5,5}^{(a)}=& \nn \frac{a}{\pi^2}\,\bigg( - \,
		\epsilon_{abcd} \,\bm{b}\,\bm{f}^a\, \bm{f}^b\, \bm{e}^c\,\bm{e}^d    - 
		\,\epsilon_{abcd} \,\bm{\bar{\tilde{\psi}}}\, \Gamma^a \,\bm{\tilde{\psi}} \,\bm{f}^b \,\bm{e}^c\, \bm{e}^d  -2\,i\,\bm{\bar{\tilde{\psi}}}\,\bm{\psi}\,\bm{a}\,\bm{f}^a\,\bm{e}_a+ \\ & \nn+
		2\, \bm{\bar{\tilde{\psi}}} \,\Gamma_{ab} \,\Gamma_5\, \bm{\psi} \,\bm{b}\,\bm{f}^a\,\bm{e}^b  +
		2\,i\,\,\bm{\bar{\tilde{\psi}}} \,\Gamma_{ab}\,\bm{\psi}\,\bm{a} \, \bm{f}^a\, \bm{e}^b  +\frac{3}{2}\,\bm{\bar{\tilde{\psi}}}\,\Gamma^a\,\bm{\tilde{\psi}}\,\bm{\bar{\psi}}\,\Gamma_a\,\bm{\psi}\,\bm{a}+\\ & \nn +\frac{1}{4}\,\bm{\bar{\tilde{\psi}}}\,\Gamma^{ab}\,\bm{\tilde{\psi}}\,\bm{\bar{\psi}}\,\Gamma_{ab}\,\bm{\psi}\,\bm{a} - 4\,\bm{\bar{\tilde{\psi}}}\,\Gamma_a\,\bm{\tilde{\psi}}\,\bm{\bar{\tilde{\psi}}}\,\Gamma_5\,\bm{\psi}\,\bm{e}^a+ \\ & -\frac{1}{8}\,\epsilon_{abcd}\,\bm{\bar{\tilde{\psi}}}\,\Gamma^{ab}\,\Gamma_5\,\bm{\tilde{\psi}}\,\bm{\bar{\psi}}\,\Gamma^{cd}\,\Gamma_5\,\bm{\psi}\,\bm{b}\bigg)
	\end{align}
\end{subequations}
For the $c$-anomaly 
\begin{equation}
H_6^{(c)}  = \delta\, Q^{(c)}_5
\end{equation}
we can choose a homogeneous representative, like in the bosonic case
\begin{equation}\label{secscg:Q5canomaly}
    \begin{split}
Q^{(c)}_5=  Q_{5,3}^{(c)} = & \frac{c}{16 \,\pi^2}\, \bigl(   
  \bm{b} \,\epsilon_{abcd} \,  \bm{\Tilde{R}}^{ab} \, \bm{\Tilde{R}}^{cd} + 2 \,  \bm{a} \,  \bm{\Tilde{R}}_{ab} \,\bm{\Tilde{R}}^{ab} - 32 \, i \, \bm{a} \,  \bm{\bar{\Tilde{\rho}}}\, \bm{\rho} - 
 32  \, \bm{b} \,\bm{\bar{\Tilde{\rho}}}\, \Gamma_5\, \bm{\rho}  \\ & 
   - 8 \,  \bm{a} \, \bm{\Tilde{F}}_R^2 - 
 32 \, i \,  \bm{\bar{\Tilde{\psi}}}  \,  \bm{\rho} \, \bm{\Tilde{F}}_R + 
 8 \,  \bm{b} \,\bm{\Tilde{F}}_R  \,\bm{\Tilde{F}}_W  - 
 8  \,  \bm{\Tilde{\psi}} \, \Gamma^{a b} \,  \Gamma_5 \, \bm{\Tilde{R}}_{ab}  \,  \bm{\rho}\bigl).
    \end{split}
\end{equation}
The anomaly density of ghost number 1 associated to this representative $Q_{5,3}^{(c)}$ is strictly BRST-invariant, not just BRST invariant modulo divergence. It the supersymmetric extension of the conformal anomaly density of $B$-type of \cite{Deser:1993yx}.

For the anomaly associated to $W^{(B)}_{6,4}$
\begin{equation}
H_{6,4}^{(B)}=c\, W_{6,4}^{(B)} = \delta\, Q^{(a=c)}_5
 \end{equation}
which describes holographic $a=c$ superconformal theories,  we can choose a representative in the degree 5 picture
\begin{align} 
Q^{(a=c)}_{5,4}=& \nn \frac{a}{\pi^2}\,\bigg(-2\,i\,\bm{\bar{\tilde{\psi}}}\,\Gamma_a\,\Gamma_5\,\bm{\tilde{\rho}}\,\bm{e}^a\,\bm{b}+2\,\bm{\bar{\tilde{\rho}}}\,\Gamma_5\,\bm{\psi}\,\bm{f}^a\,\bm{e}_a+2\,\bm{\bar{\tilde{\rho}}}\,\Gamma_{ab}\,\Gamma_5\,\bm{\psi}\,\bm{f}^a\,\bm{e}^b+ \\ & \nn -2\, \bm{\bar{\tilde{\psi}}}\,\Gamma_5\,\bm{\psi}\,\bm{a}\,\bm{\tilde{F}_R}+\bm{\bar{\tilde{\psi}}}\,\Gamma_a\,\bm{\tilde{\psi}}\,\bm{e}^a\,\bm{\tilde{F}_R}-2\,i\,\bm{\bar{\tilde{\psi}}}\,\bm{\psi}\,\bm{b}\,\bm{\tilde{F}_R}-2\,i\,\bm{\bar{\tilde{\psi}}}\,\bm{\psi}\,\bm{a}\,\bm{\tilde{F}_W}+ \\ & +2\,\bm{\bar{\tilde{\psi}}}\,\Gamma_5\,\bm{\psi}\,\bm{b}\,\bm{\tilde{F}_W}+2\,\bm{\bar{\tilde{\psi}}}\,\Gamma_5\,\bm{\psi}\,\bm{e}^a\,\bm{\tilde{T}}_a- 2\,\bm{b}\,\bm{a}\,\bm{e}^a\,\bm{\tilde{T}}_a-2\,\bm{\bar{\tilde{\psi}}}\,\Gamma_{ab}\,\Gamma_5\,\bm{\psi}\,\bm{e}^a\,\bm{\tilde{T}}^b \bigg) \\  
	Q^{(a=c)}_{5,5}=& \nn \frac{a}{\pi^2}\,\bigg( +2\,\bm{a}\,\bm{f}^a\,\bm{f}^b\,\bm{e}_a\,\bm{e}_b- 
	\epsilon_{abcd} \,\bm{b}\,\bm{f}^a\, \bm{f}^b\, \bm{e}^c\,\bm{e}^d  -2\,\bm{\bar{\tilde{\psi}}}\,\Gamma_a\,\bm{\tilde{\psi}}\,\bm{b}\,\bm{a}\,\bm{e}^a +\\ & \nn + 
	\,\epsilon_{abcd} \,\bm{\bar{\tilde{\psi}}}\, \Gamma^a \,\bm{\tilde{\psi}} \,\bm{f}^b \,\bm{e}^c\, \bm{e}^d+2\,\bm{\bar{\psi}}\,\Gamma_a\,\bm{\psi}\,\bm{a}\,\bm{b}\,\bm{f}^a-2\,\epsilon_{abcd}\,\bm{\bar{\psi}}\,\Gamma^a\,\bm{\psi}\,\bm{f}^b\,\bm{f}^c\,\bm{e}^d+ \\ & \nn -2\,\bm{\bar{\tilde{\psi}}}\,\Gamma_5\,\bm{\psi}\,\bm{b}\,\bm{f}^a\,\bm{e}_a  +2\,\bm{\bar{\tilde{\psi}}}\,\Gamma_{ab} \,\Gamma_5\,\bm{\psi}\,\bm{b}\,\bm{f}^a\,\bm{e}^b+\bm{\bar{\tilde{\psi}}}\,\Gamma_{a}\,\bm{\tilde{\psi}}\,\bm{\bar{\psi}}\,\Gamma^a\,\bm{\psi}\, \bm{a}+ \\ & - 2 \bm{\bar{\tilde{\psi}}}\,\Gamma_a\,\bm{\tilde{\psi}}\,\bm{\bar{\tilde{\psi}}}\,\Gamma_5\,\bm{\psi}\,\bm{e}^a  \bigg)
\end{align}
with vanishing $(5,3)$ component
\begin{equation}
Q^{(a=c)}_{5,3}= 0 .
\end{equation}

\subsection{The Superconformal Anomaly Densities}\label{sec:anomalydensities}
The quantum effective action is  a (non-local) functional of the independent fields of superconformal gravity
\begin{equation}
\mathscr{W}[e,b,a,\psi] = -i\,\log{Z[e,b,a,\psi]}.
\end{equation}
which defines the currents
\begin{equation}
{\mathcal{T}}_a{}^\mu = e^{-1}\,\tfrac{\delta \mathscr{W}}{\delta e_\mu{}^a},\quad
{\mathcal{B}}^\mu = e^{-1}\,\tfrac{\delta \mathscr{W}}{\delta b_\mu},\quad
{\mathcal{J}}^\mu = e^{-1}\,\tfrac{\delta \mathscr{W}}{\delta a_\mu},\quad
{\mathcal{S}}^\mu = e^{-1}\,\tfrac{\delta \mathscr{W}}{\delta \bar{\psi}_\mu}.
\end{equation}
The most general anomaly cocycle depends on all  superconformal ghosts. As shown in Subsection \ref{sec:scganomalyreprs} one can choose anomaly representatives which are independent of the Lorentz ghosts and connections. 
One can also eliminate the ghosts $\theta^a$ associated to special superconformal transformations by putting to zero the field $b$ and at the same time performing  the substitution \cite{Imbimbo:2023sph}
\begin{equation}\label{sec:fullanomalythetasubstitution}
\theta^a\to  -\tfrac{1}{2}\, e^{a\mu}\, \bigl[\partial_\mu\,\sigma+ 2\,i\,(\overline{\psi}_\mu\,\eta+\overline{\zeta}\,\Tilde{\psi}_\mu\bigr)\bigr],
\end{equation}
With this choice, the K-anomaly contributes  to the Weyl, the Q and the S anomalies.

With these renormalization prescriptions  the superconformal anomaly is described by a ghost number 1 density 
depending on the ghosts $\sigma$, $\alpha$, $\zeta$ and $\eta$ associated to Weyl, chiral, supersymmetry and special supersymmetry gauge transformations:
\begin{align} \label{eq:mathcal_A_defs}
& s\, \mathscr{W}[e,b,a,\psi]  \equiv \int\diff^4x\,e\,\bigl[\sigma\,\mathcal{A}_W + \alpha\,\mathcal{A}_R  + \bar{\zeta}\,\mathcal{A}_Q + \bar{\eta}\,\mathcal{A}_S\bigr],
\end{align}
The anomalous Ward identities take the form
\begin{equations}\label{sec5:wardidentities}
& 0= \tfrac{1}{2}\,{\mathcal{T}}_{[ab]} - \tfrac{1}{4}\,\bar{\psi}_\mu\,\Gamma_{ab}\,{\mathcal{S}}^\mu, \\
 &0= \mathcal{B}_\mu, \\
&\mathcal{A}_W = - {\mathcal{T}}_\mu{}^\mu - \tfrac{1}{2}\,\bar{\psi}_\mu\,{\mathcal{S}}^\mu,  \\
& \mathcal{A}_R= -D_\mu\,{\mathcal{J}}^\mu + \tfrac{3}{2}\,i\,\bar{\psi}_\mu\,\Gamma_5\,{\mathcal{S}}^\mu,  \\
& \mathcal{A}_Q= D_\mu\,{\mathcal{S}}^\mu - 2\,\Gamma^a\,\psi_\mu\,{\mathcal{T}}_a{}^\mu + 2\,\Gamma_5\,\tilde{\psi}_\mu\,{\mathcal{J}}^\mu,   \\
&\mathcal{A}_S = - 2\,\Gamma_5\,\psi_\mu\,{\mathcal{J}}^\mu + i\,\Gamma_\mu\,{\mathcal{S}}^\mu,
\end{equations}
The  anomaly densities $\mathcal{A}_W $, $\mathcal{A}_R$, $\mathcal{A}_Q$ and $\mathcal{A}_S$ can be read-off  
from the generalized Chern-Simons polynomials in Eqs. (\ref{secscg:Q5aanomaly}) and (\ref{secscg:Q5canomaly}). 
 To compare with existing literature, it is convenient to write  the superconformal curvatures in terms of ordinary curvatures:
\begin{equations}
& T^a =(\dd\, e^a+\omega^a{}_b\,e^b+b\,e^a)+\overline{\psi}\,\Gamma^a\,\psi=D\,e^a+\overline{\psi}\,\Gamma^a\,\psi, \\
& \Tilde{R}^{ab}={R}^{ab}(\omega)+2\,e^{[a}f^{b]}-2\,i\,\overline{\psi}\,\Gamma^{ab}\,\Tilde{\psi},\\
& \Tilde{F}_W =\dd\, b+2\,e^a\,f_a-2\,i\,\overline{\psi}\,\Tilde{\psi}, \\
& \Tilde{F}_R =\dd \,a-2\,\overline{\psi}\,\Gamma_5\,\Tilde{\psi}=F_R-2\,\overline{\psi}\,\Gamma_5\,\Tilde{\psi}, \\
& \rho=(\dd+\tfrac{1}{4}\,\omega^{ab}\,\Gamma_{ab}+\tfrac{1}{2}\,b-\tfrac{3}{2}\,i\,a\Gamma_5)\,\psi+i\,e^a\Gamma_a\Tilde{\psi}=D\,\psi+i\,e^a\,\Gamma_a\,\Tilde{\psi}, \\
&\Tilde{T}^a =(\dd f^a+\omega^a{}_b\,f^b-b\,f^a)-\overline{\Tilde{\psi}}\,\Gamma^a\,\Tilde{\psi}=D\,f^a-\overline{\Tilde{\psi}}\,\Gamma^a\,\Tilde{\psi}, \\
& \Tilde{\rho}=(\dd+\tfrac{1}{4}\,\omega^{ab}\,\Gamma_{ab}-\tfrac{1}{2}\,b+\tfrac{3}{2}\,i\,a\Gamma_5)\,\Tilde{\psi}-i\,f^a\,\Gamma_a\,{\psi}=D\,\Tilde{\psi}-i\,f^a\,\Gamma_a\,{\psi}
\end{equations}
To obtain explicit results one needs to replace $\tilde{\psi}$ and $f^a$ with their expressions in terms of the fundamental fields $e^a$, $\psi$ and $a$
\begin{equations}
	& f_{ab} \equiv  e_b{}^\mu\, f_\mu{}^a =  \nn\, -\tfrac{1}{4}\,\mathcal{R}_{ab} + \tfrac{1}{24}\,\mathcal{R}\,\eta_{ab} + \tfrac{1}{4}\,(\star\tilde{F}^R)_{ab} \,+\nonumber \\
	& \qquad  -\tfrac{1}{2}\,\bar{\psi}^c\,\Gamma_b\,\rho_{ac} 
	-\tfrac{i}{2}\,\bar{\psi}^c\,\Gamma_{ca}\,\tilde{\psi}_b + \tfrac{i}{2}\,\bar{\psi}_b\,\Gamma_{ca}\,\tilde{\psi}^c + \tfrac{i}{6}\,\eta_{ab}\,\bar{\psi}^c\,\Gamma_{cd}\,\tilde{\psi}^d \\ 
	& \tilde{\psi}_a = \tfrac{i}{2}\,\Gamma^b\,D_{[b}\psi_{a]}  -  \tfrac{i}{12}\,\Gamma_a\,\Gamma^{bc}D_{[b}\psi_{c]}.
\end{equations}
where $\mathcal{R}_{ab}$ is the Ricci tensor associated to the ordinary Riemann curvature ${R}^{ab}(\omega)$:
\begin{equation}
\mathcal{R}_{ab} \equiv   R_{\mu\rho; b}{}^{\;c}(\omega)\, e_c{}^\rho\, e_{a}{}^\mu
\end{equation}

The degree 6 picture of the  generic superconformal anomaly defines the generalized Chern-Simons polynomial  $Q_5^{(a)} + Q_5^{(c)}$
\begin{equation}\label{secad:q5ac}
H_6^{(a)}+ H_6^{(c)} =   (a-c) \,W^{(A)}_{6,3}+ c\, W^{(B)}_{6,4}= \delta (Q_5^{(a)} + Q_5^{(c)})
\end{equation}
up to $\delta$-exact $Q_4$ polynomials
\begin{equation}
Q_5^{(a)} + Q_5^{(c)}\to Q_5^{(a)} + Q_5^{(c)}+ \delta\, Q_4 .
\end{equation}
By  choosing $Q_4$ appropriately one can put to zero  the  portion of the  supersymmetry $Q$-anomaly $\mathcal{A}_Q$  which describes anomalous contributions to the correlators of the divergence of the  supercurrent in presence of another fermionic currents and a bosonic current. We call this part of the anomaly the ``cubic''  Q-anomaly.  

The representatives for $Q_5^{(a)}$ and $Q_5^{(c)}$ in Eqs.  (\ref{secscg:Q5aanomaly}) and (\ref{secscg:Q5canomaly}) both  give rise to null cubic Q-anomalies. Putting to zero the cubic Q-anomaly does not yet completely fix the freedom to add  $\delta$-exact terms to the anomaly. We considered two more choices of renormalization prescriptions which further specify the anomaly representatives:

\begin{itemize}

\item One choice is to pick $Q_5^{(a)}$  as in  Eq.  (\ref{secscg:Q5aanomaly}) and  $Q_5^{(c)}=Q_{5,3}^{(c)}$,  as in Eq. (\ref{secscg:Q5canomaly}), corresponding to a $c$-anomaly density which is strictly $s$-invariant, not just invariant up to divergences. This still leaves two   parameters $k_1$ and $k_2$ free. The cubic $\mathcal{A}_{R,W,S}$ anomalies and the quartic $\mathcal{A}_Q$ anomaly with this renormalization prescriptions are\footnote{In these and in the following formulas we renormalize the fields as \begin{align} 
				& 	\sigma\rightarrow -\sigma \\
				& \alpha \rightarrow \frac{2}{3}\alpha \qquad a \rightarrow \frac{2}{3}a \qquad F_R \rightarrow \frac{2}{3}F_R \\ 
				& \zeta \rightarrow \frac{1}{2}\zeta \qquad \psi \rightarrow \frac{1}{2}\psi \qquad D\psi \rightarrow \frac{1}{2}D\psi \\ 
				& \eta \rightarrow \frac{1}{2}\eta \qquad \tilde{\psi} \rightarrow \frac{1}{2}\tilde{\psi} \qquad D\tilde{\psi} \rightarrow \frac{1}{2}D\tilde{\psi}  
\end{align} to match with the existing literature.}:
\begin{subequations}\label{secad:acB}
	\begin{align}
		& \sigma \mathcal{A}_W=-\frac{a}{16\pi^2}\,\sigma\, E_4+\frac{c}{16\pi^2}\,\sigma\, W^{\alpha\beta\gamma\delta}W_{\alpha\beta\gamma\delta}-\frac{c}{6\pi^2}\,\sigma \,F_R^{\alpha\beta}F_R\,_{\alpha\beta}+ O(\psi^2)\\
		& \alpha \mathcal{A}_R=\frac{(5 a-3c)}{54\pi^2}\,\alpha\, \epsilon_{\alpha\beta\gamma\delta}\,F_R^{\alpha\beta}F_R^{\gamma\delta}-\frac{(a-c)}{48\pi^2}\,\alpha\,\epsilon_{\alpha\beta\gamma\delta}\,W^{\alpha\beta\lambda\mu}\,W^{\gamma\delta}\,_{\lambda\mu} + O(\psi^2)   \\
		& \nonumber \bar{\eta}\mathcal{A}_S=-\frac{i((5+ k_1) a- c)}{24\pi^2}\,\bar{\eta}\,\epsilon_{\alpha\beta\gamma\delta}\,D^{\alpha}\psi^{\beta}\,F_R^{\gamma\delta}  +\frac{c}{12\pi^2}\,\bar{\eta}\,\Gamma_5\,D^{[\alpha}\,\psi^{\beta]}(\delta_{\mu}\,^{\alpha}\delta_{\nu}\,^{\beta}+\\ & \qquad \nn + \Gamma_{\beta\nu}\delta_{\beta}\,^{\mu})F_R^{\mu\nu}  -\frac{i a}{4\pi^2}\,\bar{\eta}\,\Gamma_{\alpha\gamma}\,D^{[\beta}\,\psi^{\gamma]}\,S^{\alpha}\,_{\beta} + \frac{i a}{48\pi^2}\,\bar{\eta}\,\Gamma_{\alpha\beta}\,D^{\alpha}\,\psi^{\beta}\,R\\ & \qquad \nn +\frac{i( a- c)}{16\pi^2}\,\bar{\eta}\,\Gamma^{\gamma\delta}\,D^{\alpha}\,\psi^{\beta}\,W_{\gamma\delta\alpha\beta}   +k_2\, a\, (\bar{\eta}\,\Gamma_5D^{\alpha}\,\psi^{\beta}\,F_R^{\alpha\beta}+\\ & \qquad + \bar{\eta}\,\Gamma_{\beta\gamma}\,\Gamma_5\,D^{[\alpha}\,\psi^{\beta]}\,F_R\,_{\alpha}\,^{\gamma} -\frac{1}{4}\,\epsilon_{\alpha\beta\gamma\delta}\,\bar{\eta}\,\Gamma^{\beta}\,\Gamma_5\,D^{\gamma}\,\psi^{\delta}\,a^{\alpha}) \\
		& \nonumber \bar{\zeta}\mathcal{A}_Q= -\frac{(5 a+3c)}{108\pi^2}\,\bar{\zeta}\,\Gamma_{\alpha}  \,D_{\beta}\,\psi_{\gamma}\,a^{\alpha}\,F_R^{\beta\gamma}+\frac{c}{36\pi^2}\,\bar{\zeta}\,\Gamma^{\beta}\,D_{[\beta}\,\psi_{\gamma]}\,a_{\alpha}\,F_R^{\alpha\gamma}+ \\ &  \qquad \nonumber +\frac{(10 a-3c)}{108\pi^2}\,\bar{\zeta}\,\Gamma_{\beta}\,D_{[\alpha}\,\psi_{\gamma]}\ a^{\alpha}F_R^{\beta\gamma}  +i \frac{(10a-3c)}{216\pi^2}\,\bar{\zeta}\,\epsilon_{\beta\gamma\delta\lambda}\,\Gamma_{\alpha}\,\Gamma_5\,D^{\beta}\,\psi^{\gamma}\,a^{\alpha}F_R^{\delta\lambda}+ \\ & \qquad \nonumber +\frac{ic}{72\pi^2}\,\bar{\zeta}\,\epsilon_{\beta\gamma\delta\lambda}\,\Gamma_{\beta}\,\Gamma_5\,D^{\gamma}\,\psi^{\delta}\,a_{\alpha}F_R^{\alpha\lambda}\\ & \quad \quad \nonumber -i\frac{(20a+3c)}{216\pi^2}\,\bar{\zeta}\,\epsilon_{\alpha\gamma\delta\lambda}\,\Gamma_{\beta}\,\Gamma_5\,D^{\gamma}\,\psi^{\delta}\,a^{\alpha}F_R^{\beta\lambda}-\frac{a}{12\pi^2}\,\bar{\zeta}\,\epsilon_{\alpha\gamma\delta\lambda}\,\Gamma^{\beta}\,D^{\delta}\,\psi^{\lambda}\,a^{\alpha}\,S_{\beta}\,^{\gamma}\\ & \quad \quad \nonumber -\frac{ic}{18\pi^2}\,\bar{\zeta}\,\Gamma^{\beta}\,\Gamma_5\,D^{[\gamma}\,\psi^{\delta]}\,a^{\alpha}\,W_{\beta\gamma\alpha\delta}+\frac{ic}{72\pi^2}\,\bar{\zeta}\,\Gamma^{\beta}\,\Gamma_5\,D^{\gamma}\,\psi^{\delta}\,a^{\alpha}\,W_{\gamma\delta\alpha\beta}\\ & \quad \quad  -\frac{c}{72\pi^2}\,\bar{\zeta}\,\epsilon_{\gamma\delta\lambda\mu}\,\Gamma_{\beta}\,D^{\gamma}\,\psi^{\delta}\,a_{\alpha}W^{\beta\lambda\alpha\mu}+\frac{(3a-2c)}{72\pi^2}\,\bar{\zeta}\,\epsilon_{\gamma\delta\lambda\mu}\,\Gamma_{\beta}\,D^{\gamma}\,\psi^{\delta}\,a_{\alpha}W^{\lambda\mu\alpha\beta}
	\end{align}
\end{subequations}
For any value of $k_1$ and $k_2$, the  $O(\psi^2)$  part of the chiral $\mathcal{A}_R$ anomaly is non-zero. In other words, with this renormalization prescription  the correlators of  the divergence of the chiral current in presence of  2 supercurrents do not vanish. 

\item A different renormalization prescription is possible which  makes all \textbf{fermionic} contributions to  the cubic chiral anomaly $\mathcal{A}_R$ vanish. This is the renormalization prescription of  \cite{Papadimitriou:2019yug}. $Q^{(a)}_5$ in Eqs. (\ref{secscg:Q5aanomaly})  satisfies such renormalization choice, while the $Q_5^{(c)}$ which meets these conditions differs by $\delta$-exact terms from the polynomial in Eq. (\ref{secscg:Q5canomaly}).  The total cubic $\mathcal{A}_{R,W,S}$ anomalies and the quartic $\mathcal{A}_Q$ anomaly with this renormalization prescriptions are:
\begin{subequations}\label{secad:acR}
	\begin{align}
		& \sigma \mathcal{A}_W=-\frac{a}{16\pi^2}\,\sigma\, E_4+\frac{c}{16\pi^2}\,\sigma\, W^{\alpha\beta\gamma\delta}\,W_{\alpha\beta\gamma\delta}-\frac{c}{6\pi^2}\,\sigma\, F_R^{\alpha\beta}F_R\,_{\alpha\beta}+ O(\psi^2)\\
		& \alpha \mathcal{A}_R=\frac{(5 a-3c)}{54\pi^2}\,\alpha \, \epsilon_{\alpha\beta\gamma\delta}\,F_R^{\alpha\beta}F_R^{\gamma\delta}-\frac{(a-c)}{48\pi^2}\,\alpha \,\epsilon_{\alpha\beta\gamma\delta}\,W^{\alpha\beta\lambda\mu}\,W^{\gamma\delta}\,_{\lambda\mu}    \\
		& \nonumber \bar{\eta}\mathcal{A}_S=-\frac{i(5a- 3c) }{24\pi^2}\,\bar{\eta}\,\epsilon_{\alpha\beta\gamma\delta}\,D^{[\alpha}\,\psi^{\beta]}\,F_R^{\gamma\delta}+\frac{c}{12\pi^2}\,\bar{\eta}\,\Gamma_5\,D^{[\alpha}\,\psi^{\beta]}\,(\delta_{\mu}\,^{\alpha}\delta_{\nu}\,^{\beta}+\Gamma_{\beta\nu}\delta_{\beta}\,^{\mu})F_R^{\mu\nu}\\ & \quad \quad \nn -\frac{i a}{4\pi^2}\,\bar{\eta}\,\Gamma_{\alpha\gamma}\,D^{[\beta}\,\psi^{\gamma]}\,S^{\alpha}\,_{\beta}  +\frac{i a}{48\pi^2}\,\bar{\eta}\,\Gamma_{\alpha\beta}\,D^{[\alpha}\,\psi^{\beta]}\,R \\ & \qquad +\frac{i( a- c)}{16\pi^2}\,\bar{\eta}\,\Gamma^{\gamma\delta}\,D^{[\alpha}\,\psi^{\beta]}\,W_{\gamma\delta\alpha\beta} \\
		& \nonumber \bar{\zeta}\mathcal{A}_Q= -\frac{(5 a-3c)}{108\pi^2}\,\bar{\zeta}\,\Gamma_{\alpha}\, D_{[\beta}\,\psi_{\gamma]}\,a^{\alpha}F_R^{\beta\gamma}+ \frac{(5 a-3c)}{54\pi^2}\,\bar{\zeta}\,\Gamma_{\beta}\,D_{[\alpha}\,\psi_{\gamma]}\,a^{\alpha}F_R^{\beta\gamma}\\ & \quad \quad \nonumber +\frac{i(5a-3c)}{108\pi^2}\,\bar{\zeta}\,\epsilon_{\beta\gamma\delta\lambda}\,\Gamma_{\alpha}\,\Gamma_5\,D^{[\beta}\,\psi^{\gamma]}\,a^{\alpha}F_R^{\delta\lambda}-\frac{i(5a-3c)}{54\pi^2}\,\bar{\zeta}\,\epsilon_{\alpha\gamma\delta\lambda}\,\Gamma_{\beta}\,\Gamma_5\,D^{[\gamma}\,\psi^{\delta]}\,a^{\alpha}F_R^{\beta\lambda}\\ & \quad \quad -\frac{(a-c)}{12\pi^2}\,\bar{\zeta}\,\epsilon_{\alpha\gamma\delta\lambda}\,\Gamma^{\beta}\,D^{[\delta}\,\psi^{\lambda]}\,a^{\alpha}S_{\beta}\,^{\gamma} +\frac{(a-c)}{24\pi^2}\,\bar{\zeta}\,\epsilon_{\gamma\delta\lambda\mu}\,\Gamma_{\beta}\,D^{[\gamma}\,\psi^{\delta]}\,a_{\alpha}W^{\lambda\mu\alpha\beta}
	\end{align}
\end{subequations}
Where 
\begin{align}
& E_4= R^{\alpha\beta\gamma\delta}R_{\alpha\beta\gamma\delta}-4R^{\alpha\beta}R_{\alpha\beta}+R^2 \\ 
& S^{\alpha\beta}=\frac{R^{\alpha\beta}}{2}-\frac{1}{12}g^{\alpha\beta}R \\
& W^{\alpha\beta\gamma\delta}=R^{\alpha\beta\gamma\delta}-S^{[\alpha[\gamma}g^{\beta]\delta]}
\end{align}
are the Euler density in 4D, the Schouten tensor and the Weyl tensor respectively. 
Let us remark that these expressions agree  with those given in \cite{Papadimitriou:2019yug} as far as  $ \mathcal{A}_W,  \mathcal{A}_R,  \mathcal{A}_S$  are concerned. However our result for the ``quartic'' anomaly $\mathcal{A}_Q$ does not agree with and --- we believe ---  corrects the result in  \cite{Papadimitriou:2019yug}  for the part which describes correlators of the divergence of the supercurrent with one extra super-current and two stress-energy tensors. 
\end{itemize}

\section{Conclusions and Open Problems}
\label{sec:conclusions}

We have successfully extended the Chern-Simons paradigm to provide a unified topological description of all 4D $\mathcal{N}=1$ superconformal anomalies, encompassing both $a$-type and $c$-type anomalies. This required several significant conceptual advances beyond the tra\-di\-tio\-nal Chern-Si\-mons framework.

We demonstrated that all 4D anomalies  ---  both Yang-Mills and superconformal  ---  can be understood through the cohomology $H_\delta(W_6)$ of the generalized BRST operator $\delta$ acting on the constraint ideal at fermion number 6. For  4D Yang-Mills theories, the constraint ideal is generated by  cubic curvature monomials, causing $H_\delta(W_6)$ to coincide with classical Chern invariants. For superconformal gravity, the larger constraint ideal includes both curvature and connection polynomials. The $a$-anomaly corresponds to the unique invariant Chern polynomial of the superconformal algebra. The $c$-anomaly emerges from non-gauge-invariant but BRST-closed polynomials mixing curvatures and connections.

This work opens several  avenues for future research.  In the Yang-Mills case, the description of anomalies through  Chern-Simons polynomials of generalized connections/curvatures of degree $d+1$ naturally fits within the holographic framework, where these generalized connections map to ordinary connections in $(d+1)$-dimensional spacetime \cite{Witten:1998qj}. However, the current holographic understanding of Weyl anomalies \cite{Henningson:1998gx},\cite{Imbimbo:1999bj} employs fundamentally different mechanisms, and to our knowledge, no holographic interpretation of supersymmetry anomalies has yet been attempted.

Our unified description of all 4D superconformal anomalies via generalized Chern-Simons polynomials suggests the existence of a corresponding unified holographic framework. A crucial step in this direction was achieved in Section~\ref{secscg:superconformalideal}, where we reformulated both superconformal constraints and non-horizontal curvatures as polynomial relations among generalized connections and curvatures.

We also found that the $AdS_5/\mathrm{CFT}_4$   $a=c$ relation emerges naturally in our Chern anomaly representation, as it corresponds to the special homogeneous  polynomials $W_{6,4}^{(B)}$. This  raises the possibility that the $a=c$ holographic result may be explained through the geometry of these generalized classes.

Another important question we leave unanswered concerns the  algebraic topological meaning of the generalized Chern classes introduced in this work. Standard Chern polynomials are well-understood as characteristic classes of principal bundles  with connections to index theory via the Atiyah-Singer theorem. We are unaware of any analogous interpretation for the generalized Chern classes we have constructed. Clarifying whether these new structures admit a similar topological or index-theoretic significance remains an intriguing direction for future research. 

Finally, while we focused on (super)conformal gravity in 4D,  our fundamental relation
\begin{equation}\label{secconclusion:fundform}
H_\delta(P_{d+1}/W_{d+1}) \cong H_\delta(W_{d+2})
\end{equation}
makes sense in arbitrary  dimension $d$. It would be  interesting to understand to which extent this same formula accounts for all anomalies in all dimensions. 

Preliminary analysis of the  6D \textbf{bosonic} conformal case\footnote{The full 6D analysis falls outside the scope of this paper.} shows that Eq. (\ref{secconclusion:fundform})  describes the type A conformal anomaly as well as one of the three known type B anomalies, in agreement with  \cite{Boulanger:2007ab}.  A natural question for future research is whether an extension exists of the conformal topological ring $P(\bm{A}, \bm{F})$ --- perhaps through enlarged algebraic structures,  incorporation of the Hodge star operator, or other geometric objects ---   to provide a unified topological characterization of all bosonic conformal anomalies in higher dimensions.   Another promising direction of investigation is to explore in concrete examples how our mechanism operates in higher dimensional theories with  extended supersymmetry.

\section*{Acknowledgments}
C.I. is deeply grateful to M. Fr\"ob for  many patient  and very helpful lessons about \textsc{FieldsX}. 

This work is  supported in part by the Italian Istituto Nazionale di Fisica Nucleare and by Research Projects, F.R.A. 2024 of the University of Genova.

\appendix

\section{\texorpdfstring{$d=4$, $\mathcal{N}=1$}{d4N1} Lie superconformal algebra}\label{appendixA}

In this Appendix  we  review our conventions for the $d=4$, $\mathcal{N}=1$ superconformal algebra. The bosonic and fermionic generators, 
the corresponding gauge fields and BRST ghosts are listed in Table \ref{Table1}.
\begin{table}[ht]
\begin{center}
\begin{tabular}{ | c | c | c c | c |} 
\hline
 Bosonic Symmetry & Generator & Gauge field & & Ghost  \\
 \hline
 Local Lorentz & $J_{ab}$ & spin connection & $\omega_\mu{}^{ab}$ & $\Omega^{ab}$ \\
 Weyl & $W$ & dilaton & $b_\mu$ & $\sigma$ \\
 $U(1)_R$ chiral R-symmetry & $R$ & $U(1)_R$-gauge field & $a_\mu$ & $\alpha$  \\
 Diffeomorphisms & $P_a$ & vierbein & $e_\mu{}^a$ & $\xi^\mu$  \\
 Special conformal & $K_a$ & conformal vierbein & $f_\mu{}^a$ & $\theta^a$  \\
 \hline
  Fermionic Symmetry & Generator & Gauge field & & Ghost  \\
 \hline
 Supersymmetry & $Q_\alpha$ & gravitino & $\psi_\mu{}^\alpha$ & $\zeta^\alpha$  \\
 Conformal supersymmetry & $S_\alpha$ & conformal gravitino & $\Tilde{\psi}_\mu{}^\alpha$ & $\eta^\alpha$ \\
 \hline
\end{tabular}
\caption{\label{Table1} $\mathfrak{su}(2,2|1)$ symmetries and generators, with their associated gauge fields and BRST ghosts.}
\end{center}
\end{table}

The (anti)-commutation relations defining the $d=4,\;\mathcal{N}=1$ superconformal algebra are:\footnote{We take spinor contractions in the $\searrow$ direction. Hence $\lambda^\alpha\chi_\alpha=-\lambda_\alpha\chi^\alpha$. E.g. $\zeta^\alpha(\Gamma^a)_{\alpha\beta}\zeta^\beta=-\zeta^\alpha(\Gamma^a)_\alpha{}^\beta\zeta_\beta=-\overline{\zeta}\,\Gamma^a\zeta$.}
\begin{equation*}
[J_{ab},J_{cd}]=\eta_{ac}\,J_{db}-\eta_{bc}\,J_{da} +\eta_{bd}\,J_{ca}-\eta_{ad}\,J_{cb}, \\[-3mm]
\end{equation*}
\begin{alignat*}{4}
& [J_{bc},P_{a}] &&= \eta_{ac}\,P_b-\eta_{ab}\,P_c, \qquad 
&& [J_{bc},K_{a}] &&= \eta_{ac}\,K_b-\eta_{ab}\,K_c, \\
 & [P_a,P_b] &&= 0, 
&& [K_a,K_b] &&= 0, \\
 & [W,P_a] &&= P_a, 
&& [W, K_a] &&= -K_a,  \\[-9mm]
\end{alignat*}
\begin{equation*}
[P_a,K_b] = 2\,(\eta_{ab}\,W+J_{ab}),\\
\end{equation*}
\begin{alignat*}{4}
 & [J_{ab},Q_\alpha] &&= \tfrac{1}{2}\,(\Gamma_{ab})_\alpha{}^\beta Q_\beta,
&& [J_{ab},S_\alpha] &&= \tfrac{1}{2}\,(\Gamma_{ab})_\alpha{}^\beta \,S_\beta, \\
 & [W,Q] &&= \tfrac{1}{2}\,Q,  
&& [W,S] &&= -\tfrac{1}{2}\,S, \\
 & [R ,Q_\alpha] &&= -\tfrac{3}{2}\,i\,(\Gamma_5)_\alpha{}^\beta\,Q_\beta,\qquad
&& [R,S_\alpha] &&= \tfrac{3}{2}\,i\,(\Gamma_5)_\alpha{}^\beta S_\beta, \\
 & [P_a,Q]&&=0,
&& [K_a,S]&&=0,\\
 & [P_a,S_\alpha]&&= i\,(\Gamma_a)_\alpha{}^\beta\,Q_\beta,
&& [K_a,Q_\alpha]&&=-i\,(\Gamma_a)_\alpha{}^\beta\,S_\beta,
\\
 & \{Q_\alpha,Q_\beta\}&&=-2\,(\Gamma^a)_{\alpha\beta}\,P_a,
&& \{S_\alpha,S_\beta\}&&=2\,(\Gamma^a)_{\alpha\beta}\,K_a, \\[-9mm]
\end{alignat*}
\begin{equation}
\{Q_\alpha,S_\beta\}=2\,i\,W\delta_{\alpha\beta}+2\,i\,(\Gamma^{ab})_{\alpha\beta}\,\tfrac{1}{2}\,J_{ab}+2\,(\Gamma_5)_{\alpha\beta}\,R.
\end{equation}
Table \ref{Table2} lists the $W$ and $R$ charges of the fields of the theory
\begin{table}[ht]
\begin{center}
\begin{tabular}{|c|c|c|}
\hline
$\bm{A}$& $W\text{-weight}$ & $R\text{-charge}$ \\
\hline
$\quad \bm{e}^a$ & $\quad\;\;\;1$ & $\quad\;\;\;0$ \\
$\quad \bm{\omega}^{ab}$ & $\quad\;\;\;0$ & $\quad\;\;\;0$ \\
$\quad \bm{b}$ & $\quad\;\;\;0$ & $\quad\;\;\;0$ \\
$\quad \bm{a}$ & $\quad\;\;\;0$ & $\quad\;\;\;0$ \\
$\quad \bm{f}^a$ & $\quad-1$ & $\quad\;\;\;0$ \\
$\quad \bm{\psi}^\alpha$ & $\quad\;\;\;\tfrac{1}{2}$ & $\quad-\tfrac{3}{2}$ \\
$\quad\bm{\tilde{\psi}}^\alpha$ & $\quad-\tfrac{1}{2}$ & $\quad\;\;\;\tfrac{3}{2}$ \\
\hline
\end{tabular}
\caption{\label{Table2} Weyl weights and R-charges of the connections.}
\end{center}
\end{table}

\section{The superconformal cocycle}\label{appendix:superconformalcocycle}
The most general closed $W_{6}$ for the superconformal theory is

\begin{align}
	W_{6,3}^{\text{closed}}	=& \nonumber	A_1 \, \bm{\bar{\tilde{\rho}}}\,\bm{\rho}\,\bm{\tilde{F}_R}+A_2\,\bm{\tilde{F}_R}\,\bm{T}^a\bm{\tilde{T}}_a+A_3 \,\epsilon_{abcd}\,\bm{\tilde{R}}^{cd}\bm{T}^a\bm{\tilde{T}}^b+ A_4 \,\bm{\tilde{F}_R}^3 + \\ &\nonumber +
	(-iA_9+A_4+iB_{13}) \,\bm{\tilde{F}_R}\, \bm{\tilde{F}}_W^2 + 
	\frac{1}{2}i(A_9-B_{13})\, \bm{\bar{\tilde{\rho}}}\,\Gamma_{ab}\,\Gamma_5\,\bm{\rho}\bm{\tilde{R}}^{ab} +\\ &\nonumber - 
	\frac{1}{8}i(A_9-B_{13})\, \bm{\tilde{F}_R}\,  \bm{\tilde{R}}_{ab}   \bm{\tilde{R}}^{ab} + 
	\frac{1}{16}i(A_9-B_{13})  \,\epsilon_{abcd}\, \bm{\tilde{F}}_W \, \bm{\tilde{R}}^{ab}  \bm{\tilde{R}}^{cd} + \\ & +
	A_9 \,\bm{\bar{\tilde{\rho}}}\,\Gamma_{a}\,\Gamma_5\,\bm{\tilde{\rho}}\,\bm{T}^a   +
	A_{10}\, \bm{\bar{\rho}}\,\Gamma_a\,\Gamma_5\,\bm{\rho}\,\bm{\tilde{T}}^a
\end{align}

\begin{align}
	W_{6,4}^{\text{closed}}&\nonumber=i(A_1+6A_9-A_{10}-6B_{13}+2i B_{14}-2B_{1}+iB_{4})\, \bm{f}^a\bm{e}_a\,\bm{\bar{\tilde{\rho}}}\,\Gamma_5\,\bm{\rho} \\ & \nonumber + 
	2\,i\,B_1\,	\bm{f}^a\bm{e}^b\, \bm{\bar{\tilde{\rho}}}\,\Gamma_{ab}\,\Gamma_5\,\bm{\rho}  +  
	2(A_9-2iA_3-B_{13}+2iB_{15}) \,\bm{f}^a\,\bm{b}\,\bm{\bar{\rho}}\,\Gamma_a\,\Gamma_5\,\bm{\rho} \\ &\nonumber+
	i\,(-2iA_1-8iA_9-2A_2+5iB_{13}+2B_{16}+B_{19}+2B_{21})\, \bm{e}^a\,\bm{b}\,\bm{\bar{\tilde{\rho}}}\,\Gamma_a\,\Gamma_5\,\bm{\tilde{\rho}}\\ &\nonumber+
	(A_1+4A_9+A_{10}-4B_{13}+2iB_{16})\, \bm{\bar{\tilde{\psi}}}\,\Gamma^a\,\bm{\tilde{\psi}}\,\bm{\bar{\rho}}\,\Gamma_a\,\Gamma_5\,\bm{\rho} \\ &\nonumber + 
	2(B_{13}-2iB_{16})\,	\bm{\bar{\tilde{\psi}}}\,\bm{\psi}\,\bm{\bar{\tilde{\rho}}}\,\Gamma_5\,\bm{\rho}-	2(B_{13}-2iB_{16})\, \bm{\bar{\tilde{\psi}}}\,\Gamma_5\,\bm{\psi}\,\bm{\bar{\tilde{\rho}}}\,\bm{\rho} \\ &\nonumber
	-(A_1+5A_9-4B_{13}+2iB_{16})\,\bm{\bar{\tilde{\psi}}}\,\Gamma^a\,\bm{\psi}\,\bm{\bar{\tilde{\rho}}}\,\Gamma_a\,\Gamma_5\,\bm{\rho} +\\ &\nonumber 
	+ (A_1+5A_9-4B_{13}+2iB_{16})\,\bm{\bar{\tilde{\psi}}}\,\Gamma^a\,\Gamma_5\,\bm{\psi}\,\bm{\bar{\tilde{\rho}}}\,\Gamma_a\,\bm{\rho} \\ & \nonumber
	+	(2A_1+8A_9-2iA_2-5B_{13}+2iB_{16}+iB_{19}+2iB_{21})\, \bm{b}\,\bm{\bar{\tilde{\psi}}}\,\bm{\rho}\,\bm{\tilde{F}}_R\\ & \nonumber+
	(iA_1+3iA_9+2A_2-B_{19}-2B_{21})\,\bm{e}^a\,\bm{\bar{\tilde{\psi}}}\,\Gamma_a\,\bm{\tilde{\rho}}\,\bm{\tilde{F}}_R\\ & \nonumber
	-(2iA_1+3iA_9+2A_2+4A_6-2B_{16}-B_{19}-2B_{21})\,\bm{a}\,\bm{\bar{\tilde{\psi}}}\,\Gamma_5\,\bm{\rho}\,\bm{\tilde{F}_R}\\ &\nonumber  -(2A_1+8A_9-2iA_2-5B_{13}+2iB_{16}+iB_{19}+2iB_{21}) \,\bm{b}\,\bm{\bar{\tilde{\rho}}}\,\bm{\psi}\,\bm{\tilde{F}}_R
	+\\ &\nonumber +(2iA_1+3iA_9+2A_2+4A_4-2B_{16}-B_{19}-2B_{21})\,\bm{a}\,\bm{\bar{\tilde{\rho}}}\,\Gamma_5\,\bm{\psi}\,\bm{\tilde{F}}_R\\ &\nonumber 
	+(\frac{1}{2}iA_1+3iA_9+2A_2-2A_4-2B_{16}-B_{19}-2B_{21})\,	\bm{\bar{\tilde{\psi}}}\,\Gamma_5\,\bm{\psi}\,\bm{\tilde{F}_R}^2 \\ & \nonumber
	-i\,(2iA_1+3iA_9+2A_2+4A_4-2B_{16}-B_{19}-2B_{21})\,	\bm{a}\,\bm{\bar{\tilde{\psi}}}\,\bm{\rho}\, \bm{\tilde{F}}_W \\ & \nonumber
	+(2iA_1+8iA_9+2A_2-5iB_{13}-2B_{16}-B_{19}-2B_{21})\,\bm{b}\,\bm{\bar{\tilde{\psi}}}\,\Gamma_5\,\bm{\rho}\, \bm{\tilde{F}}_W \\ &\nonumber
	+(A_1+3A_9-2iA_2+B_{13}+2iB_{16}+iB_{19}+2iB_{21})\,\bm{e}^a\,\bm{\bar{\tilde{\psi}}}\,\Gamma_a\,\Gamma_5\,\bm{\tilde{\rho}}\,\bm{\tilde{F}}_W \\ &\nonumber 
	+i\,(2iA_1+3iA_9+2A_2+4A_4-2B_{16}-B_{19}-2B_{21})\,\bm{a}\,\bm{\bar{\tilde{\rho}}}\,\bm{\psi}\,\bm{\tilde{F}}_W \\ & \nonumber
	-(2iA_1+8iA_9+2A_2-5iB_{13}-2B_{16}-B_{19}-2B_{21})\,\bm{b}\, \bm{\bar{\tilde{\rho}}}\,\Gamma_5\,\bm{\psi}\,\bm{\tilde{F}}_W \\ & \nonumber
	+(A_1+4A_9-A_{10}+4iA_3-4B_{13}-4iB_{15})\,\bm{f}^a\,\bm{\bar{\rho}}\,\Gamma_a\,\Gamma_5\,\bm{\psi}\,\bm{\tilde{F}}_W \\ & \nonumber -(iA_9+2A_3-iB_{13}-B_{14}-2B_{15})\,\bm{f}^a\bm{e}_a\,\bm{\tilde{F}_R}\,\bm{\tilde{F}}_W \\ &\nonumber -2(A_1+3A_9-2iA_2+2iB_{16}+iB_{19}+2iB_{21})\,\bm{\bar{\tilde{\psi}}}\,\bm{\psi}\,\bm{\tilde{F}_R}\,\bm{\tilde{F}}_W \\ &\nonumber -(\frac{3}{2}iA_1+3iA_9-2A_2-2A_4+2B_{16}+B_{19}+2B_{21})\,	\bm{\bar{\tilde{\psi}}}\,\Gamma_5\,\bm{\psi}\,\bm{\tilde{F}}_W^2 \\ & \nonumber
	+\frac{1}{4}\,(A_1+5A_9-5B_{13})\,\bm{\bar{\tilde{\psi}}}\,\Gamma_{ab}\,\bm{\psi}\,\bm{\tilde{F}_R}\,\bm{\tilde{R}}^{ab} +\\ &\nonumber +\frac{1}{4}i(A_1+5A_9-5B_{13})\, \bm{\bar{\tilde{\psi}}}\,\Gamma_{ab}\,\Gamma_5\,\bm{\psi}\,\bm{\tilde{F}}_W\,\bm{\tilde{R}}^{ab}   +B_{13}\, \bm{e}_b\,\bm{\bar{\tilde{\psi}}}\,\Gamma^a\,\Gamma_5\,\bm{\tilde{\rho}}\,\bm{\tilde{R}}_{a}\,^{b}+\\ &\nonumber +(-A_9+A_{10}+B_{13})\,\bm{f}_b\,\bm{\bar{\rho}}\,\Gamma^a\,\Gamma_5\,\bm{\psi}\,\bm{\tilde{R}}_{a}\,^{b} 	+B_4\,\bm{f}^a\,\bm{\bar{\rho}}\,\Gamma_a\,\bm{\psi}\, \bm{\tilde{F}_R}+\\ &\nonumber +B_{14}\,\bm{f}_a\bm{e}_b\,\bm{\tilde{F}_R}\,\bm{\tilde{R}}^{ab}+B_{15}\,\epsilon_{abcd}\,	\bm{f}^c\bm{e}^d\,\bm{\tilde{F}}_W\,\bm{\tilde{R}}^{ab}+B_{16}\,\epsilon_{abcd}\,\bm{e}^d\,\bm{\bar{\tilde{\psi}}}\,\Gamma^a\,\bm{\tilde{\rho}}\,\bm{\tilde{R}}^{bc} \\ & \nonumber
	+\frac{1}{2}i(A_9+A_{10}-4iA_3-B_{13})\,\epsilon_{abcd}\,\bm{f}^d\,\bm{\bar{\rho}}\,\Gamma^a\,\bm{\psi}\,\bm{\tilde{R}}^{bc}+\\ & \nonumber- \frac{1}{2}i(A_9-2iA_3-B_{13})\,\epsilon_{abce}\,\bm{f}^e\bm{e}_d\bm{\tilde{R}}^{ab}\bm{\tilde{R}}^{cd}
	+B_{19} \,\bm{a}\,\bm{\bar{\tilde{\psi}}}\,\Gamma_a\,\bm{\tilde{\rho}}\,\bm{T}^a+ \\ & \nonumber+B_{21}\, \bm{\bar{\tilde{\psi}}}\,\Gamma_a\,\bm{\tilde{\psi}}\,\bm{\tilde{F}_R}\,\bm{T}^a+B_{22}\, \bm{f}_a\,\bm{\bar{\tilde{\psi}}}\,\Gamma_5\,\bm{\rho}\,\bm{T}^a-2i B_{23} \,\bm{f}^b\,\bm{\bar{\tilde{\psi}}}\,\Gamma_{ab}\,\Gamma_5\,\bm{\rho}\,\bm{T}^a  \\ & \nonumber
	-(2A_1+8A_9-2iA_2-5B_{13}+2iB_{16}+iB_{19}+2iB_{21})\, \bm{b}\,\bm{\bar{\tilde{\psi}}}\,\Gamma_a\,\Gamma_5\,\bm{\tilde{\rho}}\,\bm{T}^a \\ &\nonumber 
	+i\,(A_1+4A_9+A_{10}-6B_{13}-2iB_{14}+2B_{1}-iB_{4}+iB_{22})\,\bm{f}_a\,\bm{\bar{\tilde{\rho}}}\,\Gamma_5\,\bm{\psi}\bm{T}^a\\ & \nonumber
	-2\,i\,(B_{13}+B_{1}-B_{23})\,\bm{f}^b\,\bm{\bar{\tilde{\rho}}}\,\Gamma_{ab}\,\Gamma_5\,\bm{\psi}\,\bm{T}^a+B_{26}\,\bm{f}_a\,\bm{b}\,\bm{\tilde{F}_R}\,\bm{T}^a \\ & \nonumber
	+(3A_9-A_2+2A_3+4A_4+3iB_{13}-B_{14}-2B_{15}-B_{26})\,\bm{f}_a\,\bm{a}\,\bm{\tilde{F}}_W\,\bm{T}^a\\ & \nonumber
	+(-iA_9+A_2+iB_{13}+B_{14})\,\bm{f}_b\,\bm{a}\,\bm{\tilde{R}}_{a}\,^{b}\bm{T}^a+\\ & \nonumber
	+(-\frac{i}{2}A_9-A_3+\frac{i}{2}B_{13}+B_{15})\,\epsilon_{abcd}\,	\bm{f}^d\,\bm{b}\,\bm{\tilde{R}}^{bc}\bm{T}^a+ \\ & \nonumber+	\frac{1}{4}(-2iA_9-4A_3+iB_{13}-2B_{16})\,\epsilon_{abcd}\,\bm{\bar{\tilde{\psi}}}\,\Gamma^a\,\bm{\tilde{\psi}}\,\bm{\tilde{R}}^{cd}\bm{T}^b \\ &\nonumber +
	2\,(-iA_9+A_2+iB_{13}+B_{14})\, \bm{a}\,\bm{\bar{\rho}}\,\Gamma_a\,\bm{\psi}\,\bm{\tilde{T}}^a+ \\ &\nonumber + 2(A_9-2iA_3-B_{13}+2iB_{15})\, \bm{b}\, \bm{\bar{\rho}}\,\Gamma_a\,\Gamma_5\,\bm{\psi}\,\bm{\tilde{T}}^a \\ &\nonumber-\frac{i}{2}(A_1+2A_9+3A_{10}-2B_{13}+2iB_{14}+iB_{4})\, \bm{\bar{\psi}}\,\Gamma_a\,\bm{\psi}\,\bm{\tilde{F}}_R \,\bm{\tilde{T}}^a \\ & \nonumber
	-(2iA_1+7iA_9+2A_2-2A-3-4iB_{13}+2B_{15}-2B_{16}-B_{19} \\ & \nonumber\qquad -2B_{21}+B_{26})\,	\bm{a}\,\bm{b}\,\bm{T}^a\bm{\tilde{T}}_a \\ &\nonumber+ 
	\frac{1}{2}\,(3iA_1+14iA_9+5iA_{10}-4A_2-14iB_{13}+2B_{14}+4iB_{1}+B_{4} \\ &\nonumber \qquad -2\,B_{22})	\,\bm{\bar{\tilde{\psi}}}\,\Gamma_5\,\bm{\psi}\,\bm{T}^a\bm{\tilde{T}}_a \\ &\nonumber -\frac{1}{2}i(3A_1+14A_9+A_{10}-14iB_{13}-2iB_{14}4B_{1}-iB_{4})\,
	\bm{e}_a\, \bm{\bar{\tilde{\psi}}}\,\Gamma_5\,\bm{\rho}\,\bm{\tilde{T}}^a \\ & \nonumber	
	+ \frac{1}{2}\,i\,(A_1+6A_9-A_{10}-6B_{13}+2iB_{14}-4B_1+iB_4)\,	 \bm{e}^b\, \bm{\bar{\tilde{\psi}}}\,\Gamma_{ab}\,\Gamma_5\,\bm{\rho}\,\bm{\tilde{T}}^a \\ & \nonumber
	- \frac{1}{2}\,i\,(A_1+2A_9+3A_{10}-2B_{13}+2iB_{14}+iB_4)	\,\bm{e}_a \,\bm{\bar{\tilde{\rho}}}\,\Gamma_5\, \bm{\psi}\,\bm{\tilde{T}}^a \\ &\nonumber 
	+ \frac{1}{2}\,i\,(A_1+2A_9+3A_{10}-2B_{13}+2iB_{14}+iB_4)\,\bm{e}^b\, \bm{\bar{\tilde{\rho}}}\,\Gamma_{ab}\,\Gamma_5\, \bm{\psi}\,\bm{\tilde{T}}^a \\ & \nonumber
	+(2iA_1+7iA_9+2A_2-2A_3-4iB_{13}+2B_{15}-2B_{16}-B_{19}-2B_{21})\,\bm{e}_a\,\bm{b}\,\bm{\tilde{F}_R}\,\bm{\tilde{T}}^a \\ &\nonumber -(2iA_1+4iA_9+A_2+4A_4-iB_{13}-B_{14}-2B_{15}-B_{19}-2B_{21})\,	\bm{e}_a\,\bm{a}\,\bm{\tilde{F}_W}\,\bm{\tilde{T}}^a \\ & \nonumber
	+i\,(A_9+iA_2-B_{13}+iB_{14}) \, \bm{a}\,\bm{e}_b\bm{\tilde{R}}_{a}\,^{b}\bm{\tilde{T}}^a+B_{40}\,\epsilon_{abcd}\,\bm{f}^c\bm{e}^d\bm{T}^a\bm{\tilde{T}}^b \\ & \nonumber
	+	(-\frac{i}{2}A_9-A_3+\frac{i}{2}B_{13}+B_{15}) \,\epsilon_{abcd}\,\bm{e}^d\,\bm{b}\,\bm{\tilde{R}}^{bc}\bm{\tilde{T}}^a \\ & 
	+\frac{i}{2}\,(A_1+6A_9+3A_{10}-8iA_{3}-6B_{13}-4B_{1}+iB_{4}+4B_{23})\, \bm{\bar{\tilde{\psi}}}\,\Gamma_{ab}\,\Gamma_5\,\bm{\psi}\,\bm{T}^a\bm{\tilde{T}}^b 
\end{align}

\begin{align}	
	W_{6,5}^{\text{closed}}	=&  \nonumber
	(-\frac{3}{2}iA_1-5iA_9-\frac{i}{2}A_{10}-2A_2+2iB_{13}-B_{14}+2B_{16}-2iB_{1}+B_{19} \\ & \nonumber\qquad -\frac{B_4}{2}+2B_{21})\,\epsilon_{abcd}\,\bm{f}^b\bm{e}^c\bm{e}^d\,\bm{\bar{\tilde{\psi}}}\,\Gamma^a\,\bm{\tilde{\rho}}+\\ &\nonumber 
	-2\,i\,(-\frac{3}{2}iA_1-5iA_9-\frac{i}{2}A_{10}-2A_2+2iB_{13}-B_{14}+2B_{16}-2iB_{1}+B_{19} \\ &\nonumber \qquad  -\frac{B_4}{2}+2B_{21})\, \bm{f}^b\bm{e}_a\bm{e}_b\bm{\bar{\tilde{\psi}}}\Gamma^a\Gamma_5 \bm{\tilde{\rho}}+ \\ & \nonumber
	-2\,(-2iA_1-7iA_9-2A_{2}+2A_3+4iB_{13}-2B_{15}+2B_{16}+B_{19}\\ & \nonumber\qquad+2B_{21}-B_{26})\bm{f}_a\bm{a}\,\bm{b}\,\bm{\bar{\rho}}\, \Gamma^a\,\bm{\psi} \\ & \nonumber
	-\frac{1}{2}\,i\,(A_1+2A_9-A_{10}+8iA_{3}-2B_{13}+2iB_{14}-4B_1+iB_4\\ &\nonumber \qquad-2iB_{40})\,\epsilon_{abcd} \,\bm{f}^b\bm{f}^c\bm{e}^d\,\bm{\bar{\rho}}\,\Gamma^a\,\bm{\psi}\\ & \nonumber
	- (4iA_1+12iA_9+4A_2-8A_3-6iB_{13}+8B_{15}-2B_{19}\\ &\nonumber \qquad-4B_{21})\,\bm{f}^a\bm{e}_a\,\bm{b}\,\bm{\bar{\tilde{\rho}}}\,\Gamma_5\,\bm{\psi}\\ &\nonumber
	-2\,(-2iA_1-6iA_9-2A_2+4A_3+3iB_{13}-4B_{15}+2B_{16}+B_{19}\\ &\nonumber \qquad+2B_{21})\,\bm{f}^a\bm{e}^b\bm{b}\,\bm{\bar{\tilde{\rho}}}\,\Gamma_{ab}\,\Gamma_5\,\bm{\psi} \\ & \nonumber
	+\frac{1}{2}(-2iA_9-2iA_{10}-8A_3+6iB_{13}-6B_{14}-4B_{15}-8B_{16}-2B_{4}\\ & \nonumber\qquad-6iB_{23}+B_{22})\, \bm{f}^a\,\bm{\bar{\tilde{\psi}}}\,\Gamma_5\,\bm{\rho}\,\bm{\bar{\psi}}\, \Gamma_a\,\bm{\psi} \\ & \nonumber
	-(A_1+2A_9-A_{10}+8iA_3-2B_{13}+2iB_{14}-8iB_{15}-4B_{1}\\ & \nonumber\qquad+iB_{4})\,\bm{f}^a\bm{f}^b\bm{e}_b\,\bm{\bar{\rho}}\,\Gamma_a\,\Gamma_5\,\bm{\psi}\\ & \nonumber
	+\frac{1}{2}\,(-2iA_1-6iA_9+8A_3+6iB_{13}-2B_{14}-4B_{15}+2iB_{23}\\ &\nonumber \qquad+B_{22})\,\bm{f}^b\,\bm{\bar{\tilde{\psi}}}\,\Gamma_{ab}\,\Gamma_5\,\bm{\rho}\,\bm{\bar{\psi}}\,\Gamma^a\,\bm{\psi} \\ & \nonumber
	-(-2iA_9+2A_2+2iB_{13}+2B_{14}-B_{19})\,\bm{a}\,\bm{\bar{\tilde{\psi}}}\,\Gamma^a\,\bm{\tilde{\psi}}\,\bm{\bar{\rho}}\,\Gamma_a\,\bm{\psi} \\ &\nonumber 
	-(\frac{5i}{2}A_1+9iA_9+\frac{3i}{2}A_{10}+2A_2-6iB_{13}-B_{14}-6B_{16}-B_{19}\\ &\nonumber \qquad-\frac{1}{2}B_{4}-2B_{21})\,\bm{e}_a\,\bm{\bar{\tilde{\psi}}}\,\Gamma^a \bm{\tilde{\psi}}\, \bm{\bar{\tilde{\rho}}}\,\Gamma_5\,\bm{\psi} \\ & \nonumber
	+(\frac{5i}{2}A_1+9iA_9+\frac{3i}{2}A_{10}+2A_2-6iB_{13}-B_{14}-6B_{16}-B_{19}\\ &\nonumber \qquad-\frac{1}{2}B_{4}-2B_{21})\,\bm{e}^b \, \bm{\bar{\tilde{\psi}}}\,\Gamma^a \bm{\tilde{\psi}}\, \bm{\bar{\tilde{\rho}}}\,\Gamma_{ab}\,\Gamma_5\,\bm{\psi} \\ & \nonumber
	+(-iA_1-5iA_9-2iA_{10}-4A_3+6iB_{13}-B_{14}+2B_{15}+2B_{16}\\ &\nonumber \qquad-iB_{23}+\frac{1}{2}B_{22})\,\bm{f}^b\,\bm{\bar{\tilde{\psi}}}\,\Gamma^{a}\,\Gamma_5\,\bm{\rho}\,\bm{\bar{\psi}} \,\Gamma_{ab}\,\bm{\psi} \\ &\nonumber 
	-(-iA_1-5iA_9-2iA_{10}-4A_3+6iB_{13}-B_{14}+2B_{15}+2B_{16}\\ &\nonumber \qquad-iB_{23}+\frac{1}{2}B_{22})\,\bm{f}^b\,\bm{\bar{\tilde{\psi}}}\,\Gamma^{a}\,\bm{\rho}\,\bm{\bar{\psi}}\, \Gamma_{ab}\,\Gamma_5\,\bm{\psi} \\ &\nonumber 
	+i\,(2iA_1+6iA_9+2A-2-4A_3-3iB_{13}+4B_{15}-2B_{16}-B_{19}\\ &\nonumber \qquad-2B_{21})\,\bm{b}\, \bm{\bar{\tilde{\psi}}}\,\Gamma^a\,\bm{\tilde{\psi}}\,\bm{\bar{\rho}}\, \Gamma_a\,\Gamma_5\,\bm{\psi} \\ & \nonumber
	-(2iA_1+6iA_9+2A-2-4A_3-3iB_{13}+4B_{15}-2B_{16}-B_{19}\\ &\nonumber \qquad-2B_{21})\,\bm{f}^a\bm{f}^b\bm{e}_a\bm{e}_b\,\bm{\tilde{F}}_R\\ & \nonumber
	+\frac{1}{2}(2iA_1+6iA_9+2A-2-4A_3-3iB_{13}+4B_{15}-2B_{16}-B_{19}\\ &\nonumber \qquad -2B_{21})\,\bm{e}_a\,\bm{b}\,\bm{\bar{\tilde{\psi}}}\,\Gamma^a\,\bm{\tilde{\psi}}\, \bm{\tilde{F}_R}\\ & \nonumber
	+\frac{1}{2}(-3A_1-12A_9-3A_{10}+4iA_2+8iA_3+6_{13}+2iB_{14}-8iB_{15}-8iB_{16} \\ & \nonumber\qquad  -6B_1-2iB_{19}+iB_4-4iB_{21})\,\bm{f}^a\bm{e}_a\,\bm{\bar{\tilde{\psi}}}\,\bm{\psi}\,\bm{\tilde{F}_R}\\ & \nonumber
	+2\,(2iA_1+6iA_9+2A-2-4A_3-3iB_{13}+4B_{15}-2B_{16}-B_{19}\\ &\nonumber \qquad-2B_{21})\,\bm{f}^a\,\bm{b}\,\bm{\bar{\psi}}\,\Gamma_a\,\bm{\psi}\,\bm{\tilde{F}_R}\\ & \nonumber
	+(-A_1-3A_9+2iA_2+2iB_{14}-3B_1-iB_{19}+iB_{4}-2iB_{21})\, \bm{f}^a\bm{e}^b\, \bm{\bar{\tilde{\psi}}}\,\Gamma_{ab}\,\bm{\psi}\,\bm{\tilde{F}}_R\\ &\nonumber 
	+(-\frac{5i}{2}A_1-9iA_9-\frac{3i}{2}A_{10}-2A_2+6iB_{13}+B_{14}+6B_{16}+B_{19}+\frac{1}{2}B_4\\ &\nonumber \qquad+2B_{21})\, \bm{\bar{\tilde{\psi}}}\,\Gamma_a\,\bm{\tilde{\psi}}\,\bm{\bar{\psi}}\, \Gamma^a\, \bm{\psi}\,\bm{\tilde{F}}_R\\ & \nonumber
	+(-2iA_1-7iA_9-2A_2+2A_3+4iB_{13}-2B_{15}+2B_{16}\\ & \nonumber\qquad+B_{19}+2B_{21}-B_{26})\,\bm{f}^a\,\bm{a}\,\bm{e}_a\,\bm{b}\, \bm{\tilde{F}}_W \\ & \nonumber
	+\frac{1}{2}(2iA_1+6iA_9+2A_2-4A_3-3iB_{13}-2B_{16}\\ & \nonumber\qquad-B_{19}-2B_{21}+B_{40})\,\epsilon_{abcd}\,\bm{f}^a\bm{f}^b\bm{e}^c\bm{e}^d\,\bm{\tilde{F}}_W \\ & \nonumber
	+\frac{1}{2}(2iA_9-2A_2-2iB_{13}-2B_{14}+B_{19})\,\bm{a}\,\bm{e}^a\,\bm{\bar{\tilde{\psi}}}\,\Gamma_a\,\bm{\tilde{\psi}}\,\bm{\tilde{F}}_W \\ & \nonumber
	+(3iA_1-11iA_9-2iA_{10}-2A_2+9iB_{13}-4B_{14}+2B_{16}+B_{19}-B_{4}\\ & \nonumber\qquad+2B_{21}+B_{22})\,	\bm{f}^a\bm{e}_a\,\bm{\bar{\tilde{\psi}}}\,\Gamma_5\,\bm{\psi}\, \bm{\tilde{F}}_W \\ &\nonumber
	+\frac{1}{2}(-3iA_1-8iA_9+iA_{10}-4A_2+8A_3-2B_{14}+4B_{16}-6iB_1+2B_{19}-B_{4}\\ &\nonumber \qquad +4B_{21}+4iB_{23})\,	
	\bm{f}^a\bm{e}^b \,\bm{\bar{\tilde{\psi}}}\,\Gamma_{ab}\,\Gamma_5\,\bm{\psi}\,\bm{\tilde{F}}_W \\ &\nonumber
	+\frac{1}{4}i(A_1+4A_9+A_{10}-4B_{13}+2iB_{14}+4iB_{16}-2B_1+iB_{4})\,	
	\bm{f}^c\bm{e}_c\,\bm{\bar{\tilde{\psi}}}\,\Gamma_{ab}\,\Gamma_5\,\bm{\psi}\,\bm{\tilde{R}}^{ab} \\ &\nonumber
	+\frac{1}{2}(2iA_9-2A_2-2iB_{13}-2B_{14}+B_{19})\,	
	\bm{a}\,\bm{e}_b\,\bm{\bar{\tilde{\psi}}}\,\Gamma_a\,\bm{\tilde{\psi}}\,\bm{\tilde{R}}^{ab} \\ &\nonumber
	-i(A_9+A_{10}-4iA_3-2B_{13}-B_{1}+B_{23})\,	
	\bm{f}^b\bm{e}_c\,\bm{\bar{\tilde{\psi}}}\,\Gamma_{ab}\,\Gamma_5\,\bm{\psi}\,\bm{\tilde{R}}^{ac} \\ & \nonumber
	-\frac{1}{2}i(A_1+4A_9+A_{10}-4B_{13}+2iB_{14}+4iB_{16}-2B_1+iB_{4})\,	
	\bm{f}_c\bm{e}^b\,\bm{\bar{\tilde{\psi}}}\,\Gamma_{ab}\,\Gamma_5\,\bm{\psi}\,\bm{\tilde{R}}^{ac} \\ &\nonumber
	+(-2iA_1-7iA_9-2A_2+2A_3+4iB_{13}-2B_{15}+2B_{16}+B_{19}\\ &\nonumber \qquad+2B_{21}-B_{26})\,
	\bm{f}_a\,\bm{a}\,\bm{e}_b\,\bm{b}\,\bm{\tilde{R}}^{ab} \\ &\nonumber
	+\frac{1}{2}(-4B_{15}+B_{40})\,\epsilon_{acde}\,
	\bm{f}^c\bm{f}^d\bm{e}^b\bm{e}^e \bm{\tilde{R}}^{ab}+ \\ & \nonumber
	+\frac{1}{2}\,(-A_9-A_{10}+4iA_3+B_{13}-4iB_{15}-2iB_{16}-B_1)\,\epsilon_{abcd}\,
	\bm{f}^c\bm{e}^d \,\bm{\bar{\tilde{\psi}}}\,\bm{\psi}\,\bm{\tilde{R}}^{ab} \\ &\nonumber
	-\frac{1}{2}\,i\,(3A_1+12A_9+3A_{10}-14B_{13}-6iB_{14}+2B_1-iB_{4}+2iB_{22})\,	
	\bm{f}_a\bm{e}_b \,\bm{\bar{\tilde{\psi}}}\,\Gamma_5\,\bm{\psi}\,\bm{\tilde{R}}^{ab} \\ &\nonumber
	+\frac{1}{4}(2iA_1+6iA_9+2A_2-4A_5-3iB_{13}+4B_{15}-2B_{16}-B_{19}\\ &\nonumber \qquad-2B_{21})	\,\epsilon_{abcd}\,
	\bm{e}^d\,\bm{b}\,\bm{\bar{\tilde{\psi}}}\,\Gamma^a\,\bm{\tilde{\psi}}\, \bm{\tilde{R}}^{bc}\\ &\nonumber
	+(-2iA_1-7iA_9-2A_2+2A_3+4iB_{13}-2B_{15}+2B_{16}+B_{19}\\ & \nonumber\qquad+2B_{21}-B_{26})\,	
	\bm{a}\,\bm{b}\,\bm{\bar{\tilde{\psi}}}\,\Gamma^a\,\bm{\tilde{\psi}}\,\bm{T}_a \\ &\nonumber
	+\frac{3}{4}\,i\,(4A_1+18A_9+4A_{10}-4iA_3-17B_{13}+6iB_{16}+2iB_{21}\\ &\nonumber \qquad+2B_{23}+iB_{22})\,	
	\bm{\bar{\tilde{\psi}}}\,\Gamma_a \,\bm{\tilde{\psi}}\,\bm{\bar{\tilde{\psi}}}\,\Gamma_5 \,\bm{\psi} \bm{T}^a \\ & \nonumber
	-2\,(-2iA_1-7iA_9-2A_2+2A_3+4iB_{13}-2B_{15}+2B_{16}+B_{19}\\ & \nonumber\qquad+2B_{21}-B_{26})\,
	\bm{f}_a\bm{f}^b\,\bm{a}\,\bm{e}_b\bm{T}^a \\ &\nonumber
	+(2iA_1+6iA_9+2A_2-4A_5-3iB_{13}+4B_{15}-2B_{16}-B_{19}\\ &\nonumber \qquad-2B_{21})\,	\epsilon_{abcd}\,
	\bm{f}^b\bm{f}^c\bm{e}^d\,\bm{b}\,\bm{T}^a \\ &\nonumber
	+(-4A_1-12A_9+6iA_2-4iA_3+6B_{13}+2iB_{14}+4iB_{15}-4iB_{16}\\ & \nonumber\qquad-3iB_{19}-4iB_{21}+2iB_{26})\,
	\bm{f}_a\,\bm{a}\,\bm{\bar{\tilde{\psi}}}\,\bm{\psi}\,\bm{T}^a \\ &\nonumber
	+(2A_9+2iA_2-2B_{13}+2iB_{14}-iB_{19})\,
	\bm{f}^b\,\bm{a}\, \bm{\bar{\tilde{\psi}}}\,\Gamma_{ab}\,\bm{\psi}\,\bm{T}^a \\ &\nonumber
	+(-6iA_1-20iA_9-6A_2+8A_3+11iB_{13}-8B_{15}+6B_{16}+3B_{19}\\ &\nonumber \qquad+6B_{21}-2B_{26})\,	
	\bm{f}_a\,\bm{b}\, \bm{\bar{\tilde{\psi}}}\,\Gamma_5\,\bm{\psi}\,\bm{T}^a \\ &\nonumber
	-(2iA_1+6iA_9+2A_2-4A_5-3iB_{13}+4B_{15}-2B_{16}-B_{19}\\ & \nonumber\qquad-2B_{21})\,
	\bm{f}^b\,\bm{b} \,\bm{\bar{\tilde{\psi}}}\,\Gamma_{ab}\,\Gamma_5\,\bm{\psi}\,\bm{T}^a \\ &	\nonumber
	+\frac{1}{2}(3iA_1+10iA_9+iA_{10}+4A_2-2iB_{13}+2B_{14}-8B_{16}+8iB_1-2B{19}\\ & \nonumber\qquad+B_4-4B_{21}-4iB_{23}-2B_{40})\,\epsilon_{abcd}\,	\bm{f}^c\bm{e}^d\, \bm{\bar{\tilde{\psi}}}\,\Gamma^a\,\bm{\tilde{\psi}}\,\bm{T}^b\\ &\nonumber
	+\frac{1}{4}i(A_1+6A_9+3A_{10}-8iA_3-10B_{13}-8B_{1}+iB_4+8B_{23}\\ &\nonumber \qquad-2iB_{40})\,\epsilon_{abcd}\,
	\bm{f}^c\bm{f}^d\,\bm{\bar{\psi}}\,\Gamma^a\,\bm{\psi}\,\bm{T}^b \\ &\nonumber
	-\frac{1}{4}\,i\, (4A_1+18A_9+4A_{10}-4iA_3-17B_{13}+6iB_{16}+2iB_{21}+2B_{23}\\ &\nonumber \qquad+iB_{22})\,
	\bm{\bar{\tilde{\psi}}}\,\Gamma_a\, \bm{\tilde{\psi}}\,\bm{\bar{\tilde{\psi}}}\,\Gamma_5 \,\bm{\psi} \,\bm{T}^a\\ &\nonumber
	+(2iA_1+6iA_9+2A_2-4A_3-3iB_{13}+4B_{15}-2B_{16}-B_{19}\\ & \nonumber\qquad-2B_{21})\,\epsilon_{abcd}\,
	\bm{f}^b\bm{e}^c\bm{e}^d\,\bm{b}\,\bm{\tilde{T}}^a \\ &\nonumber
	+\frac{1}{4}\,i\,(5A_1+18A_9+3A_{10}-4iA_2-12B_{13}+2iB_{14}+12iB_{16}+2iB_{19}+iB_4\\ & \qquad+4\,i\,B_{21})\,\epsilon_{abcd}\,
	\bm{e}^c\bm{e}^d\,\bm{\bar{\tilde{\psi}}}\,\Gamma^a\,\bm{\tilde{\psi}}\, \bm{\tilde{T}}^b
\end{align}

\bibliographystyle{JHEP}
\bibliography{ir}

\end{document}